\documentclass[12pt]{iopart}
\usepackage{color}
\usepackage{graphics}
 \def\al{\alpha}
 \def\b{\beta}
 \def\ga{\gamma}
 \def\de{\delta}
 
 \def\ep{\varepsilon}
 \def\ze{\zeta}
 \def\th{\theta}

 \def\ph{\varphi}

 \def\df{\delta\phi}
 \def\p{\partial}
 \def\ddf{\delta\dot{\phi}}
 \def\alo{\alpha_{1}}
 \def\alt{\alpha_{2}}
 \def\tho{\theta_{1}}
 \def\tht{\theta_{2}}
 \def\bt{\bigtriangledown}
 \def\dP{\dot{\phi}}
 \def\epr{\eta^{'}}
 \def\eps{\eta^{''}}
 \def\Ga{\Gamma}
 \def\gap{{\gamma}^{'}}
 \def\dfp{{\delta \phi}^{'}}
 \def\Pig{\Pi^{\ga}}
 \def\Pif{\Pi^{\phi}}
 \def\be   {\begin{equation}}   \def\ee   {\end{equation}}
 \def\ba   {\begin{array}}      \def\ea   {\end{array}}
 \def\bea  {\begin{eqnarray}}   \def\eea  {\end{eqnarray}}
 \def\bean {\begin{eqnarray*}}  \def\eean {\end{eqnarray*}}
\begin{document}

\title{One-loop graviton corrections to the curvature perturbation from inflation}

\author{Emanuela Dimastrogiovanni}
\address{Dipartimento di Fisica "Galileo Galilei", Universit\'a di Padova 
and INFN Sezione di Padova, via F. Marzolo, 8 I-35131 Padova, Italy}
\ead{dimastro@pd.infn.it}
\author{Nicola Bartolo}
\address{Dipartimento di Fisica "Galileo Galilei", Universit\'a di Padova 
and INFN Sezione di Padova, via F. Marzolo, 8 I-35131 Padova, Italy}
\ead{nicola.bartolo@pd.infn.it}

\date{\today}

\begin{abstract}
We compute one-loop corrections to the power spectrum of the curvature perturbation in single-field slow-roll inflation arising 
from gravitons and inflaton 
interactions. The quantum corrections due to gravitons to the power spectrum of the inflaton field are computed around the time 
of horizon crossing 
and their effect on the curvature perturbation is obtained on superhorizon scales through the $\delta N$ formalism. 
We point out that one-loop corrections from the tensor modes are of the same magnitude as those coming from scalar self-interactions, 
therefore they cannot be neglected in a self-consistent calculation. 

\end{abstract}

\pacs{98.80.Cq. \hfill DFPD-A-08-08}

\maketitle

\section{Introduction}

Inflation has become the standard paradigm to understand the initial conditions for structure formation and Cosmic Microwave Background (CMB) 
anisotropies. 
In the inflationary picture, primordial density and gravity-wave fluctuations are generated from quantum fluctuations and stretched to superhorizon scales during 
an early period of accelerated expansion of the universe~\cite{lrreview}. Such fluctuations can trigger the formation of the CMB 
anisotropies taking place in a later radiation- or matter-dominated epoch. CMB observations~\cite{smoot92,bennett96,gorski96,wmap3,wmap5} 
show that the cosmological perturbations are very small,
of order $10^{-5}$ compared to the homogeneous background. This has justified the use of linear perturbation theory for a comparison with observations. There are 
however two good reasons to go beyond linear order. One is the possibility of observing some amount of non-Gaussianity 
in the CMB anisotropies, revealing fundamental details about the mechanism generating the primordial density perturbations. 
This issue is being deeply investigated at present, especially because of the 
high sensitivity to non-Gaussianity of the Planck satellite~\cite{planck} 
and its successors. Such non-Gaussianities are sourced by self-interactions in the early universe 
and become visible at the level of second- or higher-order perturbation theory~\cite{review}.
Another reason to go beyond linear order is to look at the loop corrections the self-interactions of any scalar field during an inflationay phase 
(and more interestingly of the curvature perturbation $\zeta$) can generate 
in the cosmological correlators, including the observationally interesting cases of the
two- and the three-point correlation functions (or, in Fourier space, the power spectrum and the bispectrum, respectively).         

Loop corrections may lead to interesting effects which scale like the power of the number of e-folds between horizon exit of a given mode $k$ and the end of 
inflation~\cite{Mukhanov:1996ak,Abramo:1997hu,boy1,boy2,olandesi}. 
Recently there has been some renewed interest in loop corrections to the correlators of cosmological perturbations generated during an early epoch of inflation, 
stimulated by two papers of Weinberg~\cite{weinberg1,weinberg2}. The reason is that one-loop corrections to the power spectrum of the 
curvature perturbation $\zeta$ seem to show some infra-red divergences which scale like $\ln(kL)$, where $L^{-1}$ is some infra-red comoving momentum cut-off. 
\cite{sloth,sloth2,seery1,seery2}.
\footnote{By choosing 
the smallest possible value for $L^{-1}$, $L^{-1}=a_i H$, where $a_i$ is the scale factor at the beginning of inflation, 
the logarithm becomes $\ln(kL)=\ln(a_k/a_i)$, which is proportional to the number of e-folds between the beginning of inflation and the time the mode $k$ leaves 
the horizon (when $k=a_k H$). This would have a dramatic impact for the control of cosmological perturbation theory and in particular 
for the 
comparison of inflationary predictions for the primordial density perturbations with observations.}.
However it has been discussed in Refs.~\cite{box4,box5} (see also~\cite{Enqvistetal}) that such 
potentially large corrections do not appear in quantities that are directly observable. Indeed they 
concern variations of background quantities defined within a region of comoving size $\ell$ not much larger than the present horizon $H_0^{-1}$ when one 
considers how they change on scales $L$ much larger than the size $\ell$. Such variations will be then associated to 
infrared corrections which however are not related to the local observable quantities. This means that loop corrections are under control because it is justified 
to choose for the infrared momentum cut-off the scale $\ell$ not much larger than the present horizon $H_0^{-1}$.
\footnote{In some special cases a non-linear resummation of the infra-red divergences discussed here is doable, see~\cite{RiottoSloth}.}

Independently of this issue, up to now the computations of one-loop corrections to the power spectrum of curvature perturbations have neglected any contribution 
arising from the graviton excitations (or tensor perturbation modes) which are unavoidably generated during a quasi de-Sitter phase. In particular, when 
going beyond linear order, a mixing of scalar and tensor perturbations is inevitable as well. In this paper we compute for the first time the graviton 
effects at one-loop level to the power spectrum of the curvature perturbation $\zeta$ in single-field models of slow-roll inflation. Our results 
show that such corrections are of the same order as those obtained by considering only the loops from scalar self-interactions 
(see for example~\cite{seery1,seery2}). This further strengthens the point that a full self-consistent computation of one- (or higher-order) loop corrections to 
the correlation functions of primordial curvature 
perturbations must necessarily account for the graviton contribution inside the loops. Loop corrections in general appear to be small, neverthless 
an evaluation as precise as possible is required. However notice that, interestingly enough, 
we find also for the graviton-loop contributions terms which scale like $\ln(k \ell)$.   

Most of the computation consists in deriving the quantum corrections due to gravitons to the two-point function of the inflaton field $\delta \phi$ 
around the time of horizon crossing. This is achieved using the in-in, also dubbed Schwinger-Keldysh, formalism~\cite{in-in1,in-in2,in-in3}. Their contribution to the power spectrum of the 
curvature perturbation $\zeta$ on superhorizon scales is then obtained by exploiting the $\delta N$ formalism~\cite{deltaN1,deltaN2,deltaN3,deltaN4}. 
The $\de N$ formalism allows to obtain an expression for the curvature perturbation $\ze$ (in the uniform density gauge) 
in terms of the scalar field(s) fluctuations
\be
\label{DN}
\ze(\vec{x},t)=\sum_{n}{\frac{N^{(n)}(t,t_*)}{n!}\left(\df(\vec{x},t_*)\right)^{n}}\, ,
\ee
where $N^{(n)}(t,t_*)=(\partial /\partial \phi_*)^n N(t,t_*)$ represents the derivative of the number of e-folds (between the final time $t$ and some initial time 
$t_*$) with respect to the initial value of the scalar field (in the spatially flat gauge). 
On superhorizon scales during inflation, Eq.~(\ref{DN}) represents the perturbation in the number of e-folds of expansion between widely separated Hubble 
regions due to the initial fluctuations of the scalar field $\phi_*$. 
In fact the time $t_*$ is left to be freely chosen, as long as it is set after horizon crossing, since at that time the perturbations will 
have settled to their classical value. One possibility is to take $t$ to fall a few e-foldings after horizon crossing like for example in 
\cite{byrnes,seery1,seery2}; we will adopt this choice, motivated by the possibility of simplifying our analysis and calculations (we will provide 
details about this matter in section 3).\\  

The reader might be worried about the validity of the $\delta N$ expansion~(\ref{DN}) when tensor perturbation modes $\gamma_{ij}$ 
are accounted for, since they introduce an additional degree of freedom. In 
fact the $\delta N$ formalism holds also in this case. The reason can be simply understood by considering 
that the spatial perturbed metric can be written as~\cite{SalopekBond} $h_{ij}\propto a^2(t) e^{2 \zeta} (e^{\gamma})_{ij}$ 
(where $a(t)$ is the unperturbed scale factor, see also Eq.~(\ref{smetric})) and, on 
superhorizon scales, the tensor modes can always be reabsorbed into a rescaling of the coordinates.~\footnote{
A rigorous proof that the presence of tensor modes does not alter Eq.~(\ref{DN}) can be found, e.g., in~\cite{deltaN3}, 
and amounts to consider that $(e^{\gamma})_{ij}$ has unit determinant so that it does not modify the local volume defined in terms of the 
local scale factor $a(t,{\bf x})=a(t) e^\zeta$.}

One-loop corrections to the power spectrum of $\ze$ are provided by the action of $\delta \phi$ computed up to $4$th order. 
Both corrections due to scalar 
and graviton loops will be taken into account. The tensor fluctuations will enter the calculations up to second order (higher order tensor terms are 
unnecessary to our purpose). 
A possible classification for theories of inflation is the one that distinguishes between a canonical and non-canonical form for the kinetic term in 
the Lagrangian~\cite{non canonical.1,non canonical.2}. The latter are particularly interesting for the study of non-Gaussianity \cite{gauss1,gauss2,gauss3}. 
We derive the solutions to the constraint equations for a general non-canonical Lagrangian but our loop calculations will be focused on the canonical case 
(we will cover the study of the tensor corrections in the non canonical case both for the power spectrum and the trispectrum of the 
curvature perturbations in a companion work \cite{noi}).

The paper is organized as follows: in section 2 we calculate the complete action to fourth order in the spatially flat gauge; in section 3 we study the graviton 
one-loop corrections to the power spectrum of the inflaton field perturbations; in section 4 we collect our results 
together with the one-loop corrections from scalar perturbation modes which have been already computed in Ref.~\cite{seery1,seery2}, 
to be combined in a final formula for the corrected power spectrum of the curvature perturbation; in section 4 we also draw our conclusions. Appendices are organized as follows: in Appendix A we face the problem of dealing with a lagrangian with time derivative interactions; Appendix B contains a detailed study of the slow-roll order of the different terms of the $4$th order action together with the calculation of some of the diagrams with graviton loops; Appendix C collects the complete expressions for the one-loop two-vertex diagrams.

\section{Fourth order action for a scalar field in the spatially flat gauge}

The action to third order with tensor modes was first calculated by Maldacena \cite{maldacena} with the ADM formalism \cite{adm}. Let us calculate the fourth order action using the same formalism.\\
In the spatially flat gauge the perturbed metric is

\be\label{eq2}
ds^2=-N^2dt^2+h_{ij}(dx^{i}+N^{i}dt)(dx^{j}+N^{j}dt),
\ee

\be
\label{smetric}
h_{ij}=a^{2}(t)(e^{\ga})_{ij},
\ee
where $a(t)$ is the scale factor, $\ga_{ij}$ is a tensor perturbation with $\p_{i}\ga_{ij}=\ga_{ii}=0$ (traceless and divergenceless) and 
det$(e^{\ga})_{ij}=1$. Notice that repeated lower indices are summed up with a Kronecker delta, so $\p_{i}\ga_{ij}$ stands for $\delta^{ik}\p_{i}\ga_{kj}$ and $\ga_{ij}=\delta^{ij}\ga_{ij}$.\\
The lapse and shift functions in equation (\ref{eq2}) can be written as

\bean
N=1+\al \, ,\\
N_{j}=\p_{j}\th+\b_{j}\, ,
\eean

\noindent where $\b_j$ is divergenceless and $\al$, $\th$ and $\b_{j}$ can be expanded perturbatively up to the desired order. We have exploited the gauge freedom to set two scalar and two vector modes to zero, thus leaving one scalar mode from $N$, one scalar and two vector modes from $N_{j}$ and two tensor modes (the two independent polarizations of the graviton) from $h_{ij}$ together with the inflaton field perturbation.\\

We begin with the most general expression for the action of a scalar field minimally coupled to gravity

\be
S=\frac{1}{2}\int{dtd^{3}x \sqrt{h}\left[NR^{(3)}+2NP+N^{-1}\left(E_{ij}E^{ij}-E^{2}\right)\right]},
\ee

\noindent where $P=P(X,\phi)$ is a generic function of the scalar field and of its kinetic term 
$X=\frac{1}{2}g_{\mu \nu}\p^{\mu}\phi\p^{\nu}\phi$, $R^{(3)}$ is the curvature scalar associated with the three dimensional metric $h_{ij}$,  
\bean
E_{ij}=\frac{1}{2}\left(\dot{h_{ij}}-\bt_{i}N_{j}-\bt_{j}N_{i}\right),\\
E=h^{ij}E_{ij}.
\eean

A dot indicates derivatives w.r.t. time $t$, all the spatial indices are raised and lowered with $h_{ij}$ and units of $M^{-2}_{Pl}\equiv 8 \pi G=1$ are employed.
In the canonical case $P=X-V(\phi)$, with $V=V(\phi)$ the inflaton potential.\\ 

To $4th$ order we have

\be
R^{(3)}=-\frac{1}{4}\p_{i}\ga_{al}\p_{i}\ga_{al}.
\ee

The next step consists in writing and solving the momentum and hamiltonian constraints in order to integrate out the shift and the lapse functions $N_i$ and $N$. This is done varying the action w.r.t. the shift and lapse functions respectively; the resulting equations are

\be
\bt_{i}\left[N^{-1}\left(E^{i}_{j}-\de^{i}_{j} E\right)\right]=N^{-1}P_{,X}\left[\dot{\phi}-N^{l}\p_{l}\phi\right]\p_{i}\phi,
\ee

\be
R^{(3)}+2P-4P_{,X}X-N^{-2}\left(E_{ij}E^{ij}-E^{2}\right)-2P_{,X}h^{ij}\p_{i}\phi\p_{j}\phi=0.
\ee

\subsection{Solving the constraint equations for a theory with non canonical kinetic terms}

The action to a given order $n$ only requires the constraint equations to be solved up to order $n-2$ \cite{maldacena,gauss3}. Therefore we will solve the constraints to second order in the metric and scalar field fluctuations. Let us employ the expansions

\bean
\al=\alo+\alt,\\
\b_{i}=\b_{1i}+\b_{2i},\\
\th=\tho+\tht.
\eean

The momentum constraint at first order reads

\be\label{der}
2H\p_{j}\alo-\frac{1}{2a^{2}}\p^{2}\beta_{1j}=P_{,X}\dot{\phi}\p_{j}\df\, ,
\ee
where $H=\dot a/a$ is the Hubble parameter. Eq.~(\ref{der}) can be solved to get $\alo$. Taking the derivative $\p^{j}$ of both sides of equation (\ref{der}) and using the divergenceless condition for $\b$ the result is

\be
\alo=\frac{P_{,X}\dP\df}{2H}.
\ee

Using the solution found for $\alo$, we find $\p^{2}\b_{1j}=0$, from which we can set $\b_{1j}=0$. Here $\p^{2}\equiv \delta^{ij}\p_{i}\p_{j}$, 
which we will indicate in the rest of the paper also as $\p_{i}\p_{i}$, and from now on we define $\b_{i} \equiv \b_{2i}$ for simplicity.

The momentum constraint at second order is

\bea
2H\p_{j}\alt-4H\alo\p_{j}\alo-\frac{1}{a^{2}}\p_{j}\alo\p^{2}\tho+\frac{1}{a^{2}}\p_{i}\alo\p_{i}\p_{j}\tho-\frac{1}{2}\p_{i}\alo\dot{\ga_{ij}}\nonumber\\-\frac{1}{2a^{2}}\p^{2}\b_{j}+\frac{1}{4}\dot{\ga_{ik}}\p_{i}\ga_{kj}-\frac{1}{4}\ga_{ik}\p_{i}\dot{\ga_{kj}}-\frac{1}{4}\dot{\ga_{ik}}\p_{j}\ga_{ik}+\frac{1}{2a^{2}}\p_{i}\tho\p^{2}\ga_{ij}\nonumber\\=P_{,X}\p_{j}\df\ddf+2X P_{,XX}\p_{j}\df\ddf-2X P_{,XX}\dP\alo\p_{j}\df -P_{,X}\dP\alo\p_{j}\df \nonumber \\
+P_{,X\phi }\dP\df\p_{j}\df\, .
\eea

The solutions are

\bea
\alt&=&\frac{\alo^2}{2}+\frac{1}{2Ha^2}\p^{-2}\left[\p^{2}\alo\p^{2}\tho-\p_{i}\p_{j}\alo\p_{i}\p_{j}\tho\right]+\frac{P_{,X}}{2H}\p^{-2}\Sigma \nonumber \\
&+&\frac{1}{4H}\p^{-2}\left[\dot{\ga_{ij}}\p_{i}\p_{j}\alo\right]-\frac{1}{4a^{2}H}\p^{-2}\left[\p_{i}\p_{j}\tho\p^{2}\ga_{ij}\right] \nonumber \\
&+&\frac{1}{8H}\p^{-2}\left[\p_{j}\dot{\ga_{ik}}\p_{j}\ga_{ik}\right]
+\frac{P_{,X\phi }\dP}{2H}\p^{-2}\left[(\p_{j}\df)^2+\df\p^{2}\df\right] \nonumber \\
&+&\frac{X_{}P_{,XX}}{H}\p^{-2}\left[\p^{2}\df\ddf+\p_{j}\df\p_{j}\ddf-\dP\left(\p_{j}\alo\p_{j}\df+\alo\p^{2}\df\right)\right]\, ,
\eea
where $\Sigma\equiv\p^{2}\df\ddf+\p_{j}\df\p_{j}\ddf$, and

\bea
\frac{1}{2a^{2}}\p^{2}\b_{j}&=&2H\p_{j}\alt-4H\alo\p_{j}\alo-\frac{1}{a^{2}}\p_{j}\alo\p^{2}\tho+\frac{1}{a^{2}}\p_{i}\alo\p_{i}\p_{j}\tho \nonumber \\
&-&\frac{1}{2}\p_{i}\alo\dot{\ga_{ij}} 
+\frac{1}{4}\dot{\ga_{ik}}\p_{i}\ga_{kj}-\frac{1}{4}\ga_{ik}\p_{i}\dot{\ga_{kj}}-\frac{1}{4}\dot{\ga_{ik}}\p_{j}\ga_{ik} \nonumber \\
&+&\frac{1}{2a^{2}}\p_{i}\tho\p^{2}\ga_{ij} -P_{,X}\p_{j} \df\ddf 
-2X_{}P_{,XX}\p_{j}\df\ddf\nonumber \\
&+&2X P_{,XX}\dP\alo\p_{j}\df +P_{,X}\dP\alo\p_{j}\df-P_{,X\phi }\dP\df\p_{j}\df\, .
\eea

Let us now move to the hamiltonian constraint which provides $\tho$ and $\tht$ 

\bea
\frac{4H}{a^2}\p^{2}\tho&=&-4XP_{,X}\left(\frac{\ddf}{\dP}-\alo\right)+2P_{,\phi}\df-8P_{,XX}X^2\left(\frac{\ddf}{\dP}-\alo\right)\nonumber\\&-&4 XP_{,X\phi}\df
-12H^2\alo,
\eea

\noindent at first order and

\begin{eqnarray}
\fl
-\frac{4H}{a^2}\p^{2}\tht =
(-2\alo)\Big[4XP_{,X}\frac{\ddf}{\dP}+20P_{,XX}X^{2}\frac{\ddf}{\dP}+2XP_{,X\phi}\df+8P_{,XXX}X^{3}\frac{\ddf}{\dP}\nonumber\\
+4P_{,XX\phi}X^{2}\df+\frac{4H}{a^{2}}\p^{2}\tho\Big]
-\frac{4X\left(P_{,X}+2XP_{,XX}\right)}{a^{2}\dP}\p_{i}\tho\p_{i}\df-\frac{1}{a^{4}}\left(\p^{2}\tho\right)^{2}\nonumber\\
+\frac{1}{a^{2}}\Big[-\dot{\ga_{iq}}\p_{q}\p_{i}\tho+\frac{1}{a^{2}}\left(\p_{i}\p_{j}\tho\right)^{2}\Big]+
\left(-6H^{2}+2XP_{,X}+4X^{2}P_{,XX}\right)\Big[3\alo^{2}-2\alt\Big]\nonumber\\
+4\alo^{2}\left(3X^{2}P_{,XX}+2X^{3}P_{,XXX}\right)
+\frac{\ddf^{2}}{\dP^{2}}\Big[2XP_{,X}+16X^{2}P_{,XX}+8X^{3}P_{,XXX}\Big]\nonumber\\
+\frac{\ddf\df}{\dP}\Big[4XP_{,X\phi}+8X^{2}P_{,XX\phi}\Big]
-\frac{{\left(\p_{i}\df\right)^{2}}}{a^{2}{\dP}^{2}}\Big[4X^{2}P_{,XX}-2XP_{,X}\Big]+{\df}^{2}\Big[-P_{,\phi\phi}\nonumber\\
+2XP_{,X\phi\phi}\Big]
+\frac{1}{4}\Big[\dot{\ga_{lj}}\dot{\ga_{lj}}+\frac{1}{a^{2}}\p_{a}\ga_{iq}\p_{a}\ga_{iq}\Big]-\frac{4H}{a^{2}}\ga_{ij}\p_{i}\p_{j}\tho,
\end{eqnarray}

\noindent at second order.\\

\subsection{Reduction to the canonical case}

In the canonical case, at zeroth order in perturbation theory $P=\frac{{\dot{\phi}}^{2}}{2}-V(\phi)$, so $P_{,X}=1$ and $P_{,{\phi}^{n}}=-\p^{n} V/ \p{\phi}^{n}$ with all other derivatives of $P$ being zero. The solutions above therefore reduce to

\bea
\alo=\frac{1}{2H}\dot{\phi}\df\label{K20},
\eea

\bea
\frac{4H}{a^{2}}\p^{2}\tho=-2V_{\phi}\df-2\dot{\phi}\ddf+2\alo\left(-6H^{2}+(\dot{\phi})^{2}\right),
\eea

\bea
\alt&=&\frac{{\alo}^{2}}{2}+\frac{1}{2H}\p^{-2}\Sigma+\frac{1}{2Ha^{2}}\p^{-2}\left[\p^{2}\alo\p^{2}\tho-\p_{i}\p_{j}\alo\p_{i}\p_{j}\tho\right]
 \\
&+&\frac{1}{4H}\p^{-2}\left[\dot{\ga_{ij}}\p_{i}\p_{j}\alo\right]
-\frac{1}{4Ha^{2}}\p^{-2}\left[\p_{i}\p_{j}\tho\p^{2}\ga_{ij}\right]+\frac{1}{8H}\p^{-2}\Big[\p_{j}\dot{\ga_{ik}}\p_{j}\ga_{ik}\Big]\nonumber \, ,
\eea

\bea
\frac{4H}{a^{2}}\p^{2}\tht&=&2\alo\left[2\dot{\phi}\ddf+\frac{4H}{a^{2}}\p^{2}\tho\right]+\frac{2}{a^{2}}\dot{\phi}\p_{i}\tho\p_{i}\df
-\frac{1}{a^{4}}\p_{i}\p_{j}\tho\p_{i}\p_{j}\tho \\
&+&\frac{1}{a^{4}}\left(\p^{2}\tho\right)^{2}
-\left(3\alo^{2}-2\alt\right)\left({\dot{\phi}}^{2}-6H^{2}\right)-\ddf^{2}-\frac{1}{a^{2}}\p_{i}\df\p_{i}\df\nonumber\\
&-&V_{\phi \phi}\df^{2}
-\frac{1}{4a^{2}}\p_{a}\ga_{iq}\p_{a}\ga_{iq}-\frac{1}{4}\dot{\ga_{lj}}\dot{\ga_{lj}}+\frac{1}{a^{2}}\dot{\ga_{iq}}\p_{i}\p_{q}\tho \nonumber \, ,
\eea

\bea
\frac{1}{2a^{2}}\p^{4}\b_{j}&=&\frac{1}{a^{2}}\p^{2}\alo\p_{j}\p^{2}\tho
-\frac{1}{a^{2}}\p_{m}\p_{j}\alo\p_{m}\p^{2}\tho+\frac{1}{a^{2}}\p_{m}\alo\p_{m}\p_{j}\p^{2}\tho\nonumber\\
&-&\frac{1}{a^{2}}\p_{j}\alo\p^{4}\tho-\frac{1}{a^{2}}\p_{m}\p_{j}\p_{i}\alo\p_{i}\p_{m}\tho
+\frac{1}{a^{2}}\p^{2}\p_{i}\alo\p_{i}\p_{j}\tho\nonumber\\
&-&\frac{1}{a^{2}}\p_{i}\p_{j}\alo\p_{i}\p^{2}\tho+\frac{1}{a^{2}}\p_{m}\p_{i}\alo\p_{m}\p_{i}\p_{j}\tho 
+\p_{m}\p_{j}\ddf\p_{m}\df\nonumber\\
&-&\p^{2}\ddf\p_{j}\df+\p_{j}\ddf\p^{2}\df-\p_{m}\ddf\p_{m}\p_{j}\df
-\frac{1}{2}\p^{2}\left(\dot{\ga_{ij}}\p_{i}\alo\right)\nonumber\\
&-&\frac{1}{2a^{2}}\p^{2}\left(\p_{j}\ga_{bq}\p_{q}\p_{b}\tho\right)+\frac{1}{2a^{2}}\p^{2}\left(\p^{2}\ga_{jk}\p_{k}\tho\right)
-\frac{1}{4}\p^{2}\left(\ga_{il}\p_{i}\dot{\ga_{jl}}\right)\nonumber\\
&+&\frac{1}{4}\p^{2}\left(\dot{\ga_{ik}}\p_{i}\ga_{kj}\right).\label{K21}
\eea

\noindent where $V_{\phi \phi} \equiv \p^{2} V/\p {\phi}^{2} $ and $\p^{-2}$ is the inverse of the laplacian operator. Notice that the equations (\ref{K20}) through (\ref{K21}) agree with the results obtained in \cite{sloth-seery-lidsey} if we set $\ga_{ij}$ to zero.\\

The $4th$ order contribution to the action is
\bea
S_{4}=a^{3} \int dt d^3x \Bigg[- \frac{1}{24}V_{\phi\phi\phi\phi} \df^{4}+\frac{1}{2a^{2}}\p_{(i}\b_{j)}\p_{i}\b_{j}+\frac{1}{2a^{4}}\p_{j}\th_{1}\p_{j}\df \p_{m}\th_{1}\p_{m}\df\nonumber\\
- \frac{1}{a^{2}}\ddf\left(\p_{j}\th_{2}+\b_{j}\right)\p_{j}\df+\left(\al^{2}_{1}\al_{2}-\frac{1}{2}\al_{2}^{2}\right)\left(-6H^{2}+\dot{\ph}^{2}\right)\nonumber\\
+\frac{\al_{1}}{2}\Big[-\frac{1}{3}V_{\phi\phi\phi}\df^{3}-2V_{\phi}\al_{1}^{2}\df+\al_{1}\left(-\frac{1}{a^{2}}\p_{i}\df\p_{i}\df-V_{\phi\phi}\df^{2}\right)\nonumber\\
- \frac{2}{a^{4}}\Big(\p_{i}\p_{j}\th_{2}\p_{i}\p_{j}\th_{1}-\p^{2}\th_{1}\p^{2}\th_{2}+\p_{i}\b_{j}\p_{i}\p_{j}\th_{1}\Big)
+\frac{2}{a^{2}}\Big(\dot{\phi}\Big(\p_{j}\th_{2}+\b_{j}\Big)\p_{j}\df\nonumber\\+\ddf\p_{j}\th_{1}\p_{j}\df\Big)\Big]
+\alo^{2}\Big[\frac{1}{2a^{2}}\Big(\ga_{qi}\p_{a}\p_{i}\ga_{aq}-\frac{1}{2}\p_{a}\ga_{iq}\p_{a}\ga_{iq} \Big)-\frac{1}{4}\dot{\ga_{lj}}\dot{\ga_{lj}}+\dot{\ga_{iq}}\p_{i}\p_{q}\tho\Big]
\nonumber\\-\frac{1}{a^{2}}\Big[\frac{1}{2}\ga_{ik}\ga_{kj}\p_{j}\df\p_{i}\df-\alo\ga_{ij}\p_{j}
\df\p_{i}\df+\alt\p_{i}\df\p_{i}\df\nonumber\\
-\dot{\phi}\p_{j}\df\left(\ga_{ij}\p_{i}\tht+\ga_{ij}\b_{i}+\ga_{ij}\p_{i}\tho\right)
-\p_{k}\tht\dot{\ga_{ab}}\p_{b}\ga_{ak}-\b_{k}\dot{\ga_{ab}}\p_{b}\ga_{ak}\nonumber\\-\frac{1}{2}\dot{\ga_{ab}}\b_{k}\p_{k}\ga_{ab}-\alo\Big(H\ga_{ab}\p_{a}\p_{b}\tht+\dot{\ga_{ab}}\p_{a}\p_{b}\tht+\dot{\ga_{ab}}\p_{a}\b_{b}\Big)+\frac{1}{2}\dot{\ga_{ab}}\p_{k}\ga_{ab}\p_{k}\tht\Big]\nonumber\\+\frac{1}{2a^{4}}\Big(-8\ga_{ip}\p_{i}\p_{j}\tho\p_{p}\p_{j}\tht-4\ga_{ip}\p_{i}\p{j}\tho\p_{p}\b_{j}-4\ga_{ip}\p_{p}\p_{j}\tho\p_{j}\b_{i}\nonumber\\
-\p_{q}\tho\p_{i}\ga_{jq}\p_{i}\p_{j}\tht-\p_{q}\tht\p_{i}\ga_{jq}\p_{i}\p_{j}\tho-\b_{q}\p_{i}\ga_{jq}\p_{i}\p_{j}\tho\nonumber\\
-\p_{q}\tho\p_{i}\ga_{jq}\p_{i}\b_{j}-\p_{q}\tho\p_{i}\ga_{jq}\p_{j}\b_{i}+\p_{q}\tho\p_{q}\ga_{ij}\p_{i}\p_{j}\tht\nonumber\\
+2\p_{q}\tht\p_{q}\ga_{ij}\p_{i}\p{j}\tho+2\p_{q}\tho\p_{q}\ga_{ij}\p_{i}\b_{j}+2\b_{q}\p_{q}\ga_{ij}\p_{i}\p_{j}\tho\Big)\Bigg].
\eea
Notice that the action to fourth order with gravitons was calculated in \cite{koyama} neglecting the first-order graviton contributions for a calculation of the 
(tree-level) trispectrum in general single field models of inflation. In our paper this contribution cannot be neglected since we are calculating one-loop corrections to the power spectrum of the curvature perturbations. Moreover, as we will discuss in the next section, it turns out that loops of gravitons are not suppressed by the standard slow-roll parameters, $\epsilon = -\dot{H}/H^2$ and $\eta=(1/3) (V_{\phi \phi}/H^2)$, in comparison to the loops of scalar perturbations.\\  
From this point on, we will focus on the canonical lagrangian for our calculations.

\section{Study of one-loop tensor corrections}

The power spectrum for the curvature perturbation $\zeta$ is defined by

\be
\langle\ze_{\vec{k_{1}}}(t)\ze_{\vec{k_{2}}}(t)\rangle = (2 \pi)^{3} P_{\ze}(k) \delta^{(3)}(\vec{k_{1}}+\vec{k_{2}})\, ,
\ee

We will calculate it using the $\delta N$ formula

\bea
\langle\ze_{\vec{k_{1}}}(t)\ze_{\vec{k_{2}}}(t)\rangle=\int \frac{d^{3}x_{1}}{(2 \pi)^{3}} \frac{d^{3}x_{2}}{(2 \pi)^{3}}e^{-i(\vec{k_{1}}\vec{x_{1}}+\vec{k_{2}}\vec{x_{2}})}\nonumber\\
\left\langle \left(\sum_{n}{\frac{N^{(n)}(t,t^{*})}{n!}\left(\df(\vec{x_{1}},t^{*})\right)^{n}}\right),\left(\sum_{m}{\frac{N^{(m)}(t,t^{*})}{m!}\left(\df(\vec{x_{2}},t^{*})\right)^{m}}\right)\right\rangle.
\eea

The sums can be expanded to the desired order of loop corrections to the tree level power spectrum. Up to one loop we have

\bea\label{ZZ}
\langle\zeta_{\vec{k_{1}}}(t)\zeta_{\vec{k_{2}}}(t)\rangle={N^{(1)}}^{2}\langle\df_{\vec{k_{1}}}\df_{\vec{k_{2}}}\rangle_{*}\nonumber\\
+\frac{1}{2!}N^{(1)}N^{(2)}\int d^{3}q\langle\df_{\vec{k_{1}}}\df_{\vec{q}}\df_{\vec{k_{2}}-\vec{q}}\rangle_{*}+(\vec{k_{1}}\leftrightarrow \vec{k_{2}})\nonumber\\
+\frac{1}{3!} N^{(1)}N^{(3)}\int d^{3}q d^{3}p\langle\df_{\vec{k_{1}}}\df_{\vec{q}}\df_{\vec{p}}\df_{\vec{q}+\vec{p}-\vec{k_{2}}}\rangle_{*} +(\vec{k_{1}}\leftrightarrow \vec{k_{2}})  \nonumber\\
+\frac{1}{(2!)^2} \left(N^{(2)}\right)^{2}\int d^{3}q d^{3}p\langle\df_{\vec{q}}\df_{\vec{k_{1}}-\vec{q}}\df_{\vec{p}}\df_{\vec{k_{2}}-\vec{p}}\rangle_{*}. 
\eea

\noindent where a star indicates evaluation around the time of horizon crossing. The tree-level power spectrum of the light scalar field at lowest order in slow-roll 
is given by 

\bea\label{tree}
\langle\df_{\vec{k_{1}}}\df_{\vec{k_{2}}}\rangle_*&=& (2 \pi)^{3} P(k) \delta^{(3)}(\vec{k_{1}}+\vec{k_{2}})= (2 \pi)^{3}
\frac{H_*^{2}}{2k^{3}} \delta^{(3)}(\vec{k_{1}}+\vec{k_{2}})\, ,
\eea
where $H_*$ is the Hubble parameter evaluated at horizon exit (when $k=a H$). The variance per logarithmic interval in $k$ 
is given by ${\cal P}(k)=(k^3/ 2\pi^2) P(k)$.\footnote{Using Eq.~(\ref{tree}) and the $\delta N$ formula~(\ref{ZZ}) at lowest order, one 
recovers the well known result that at tree level the power spectrum of the curvature perturbation $\zeta$ is 
${\cal P}_{\zeta}(k)=H^2_*/(m^2_{\rm Pl}\pi \epsilon_*)$, where one uses $\left( N^{(1)} \right)^2=4 \pi G/ \epsilon_*$. 
This equation confirms that for single-field models of inflation the curvature perturbation $\zeta$ is conserved on superhorizon scales. 
This is indeed true also at the non-linear level, see Refs.~\cite{SalopekBond,maldacena,KMNR,deltaN3}. On superhorizon scales also 
the tensor perturbation modes remain constant at the fully non-linear level~\cite{SalopekBond}.} Notice that the first term on the right hand side of Eq. (\ref{ZZ}) includes both the tree level and the quantum one-loop contribution; the other terms represent the classical one-loop parts. The distinction between classical and quantum loops is understood as for example in \cite{seery2}: quantum loops find their origin in the lagrangian interaction terms between the inflaton perturbations and the gravitational modes or from self-interaction of $\df$ and are present in the expextation values of $\delta \phi$ around the time of horizon crossing; 
classical loops are corrections coming from the expansion of $\zeta$ using the $\delta N$ formula. Let us study the quantum loops.\\

The power spectrum of $\df$ can be calculated in perturbation theory using the in-in formalism (see Refs. \cite{in-in1,in-in2,in-in3} or \cite{weinberg1} which also contains a detailed review). In this formalism the expectation value of a field operator $\Theta(t)$ is given by

\be
\langle\Omega|\Theta(t)|\Omega\rangle=\left\langle 0\left|\left[\bar{T}\left(e^{i {\int}^{t}_{0}H_{I}(t')dt'}\right)\right]\Theta_{I}(t)\left[T \left(e^{-i {\int}^{t}_{0}H_{I}(t')dt'}\right)\right]\right|0\right\rangle,
\ee  

\noindent where $|\Omega\rangle$ represents the vacuum of the interacting theory, $T$ and $\bar{T}$ are time-ordering and anti-time-ordering operators, the subscript $I$ indicated the fields in the interaction picture, i.e. free fields and $H_{I}$ is the interaction hamiltonian. The interaction picture has the advantage of allowing to deal with free fields only; the fields can be thus Fourier expanded in terms of quantum creation and annihilation operators

\bean
\df(\vec{x},t)=\int d^{3}k e^{i\vec{k}\vec{x}}\left[a_{\vec{k}} \df_{k}(t)+a^{+}_{-\vec{k}} \df_{k}^{*}(t)\right],\\
\ga_{ij}(\vec{x},t)=\int d^{3}k e^{i\vec{k}\vec{x}}\sum_{\lambda}{ \left[\ep_{ij}(\hat{k},\lambda)b_{\vec{k},\lambda}\ga_{k}(t)
+\ep^{*}_{ij}(-\hat{k},\lambda)b^{+}_{-\vec{k},\lambda}\ga^*_{k}(t)\right]},
\eean

\noindent where the commutators are the usual ones

\bean
\left[a_{\vec{k}},a^{+}_{\vec{k'}}\right]=(2 \pi)^{2}\de^{(3)}(\vec{k}-\vec{k'}),\\
\left[b_{\vec{k},\lambda},b^{+}_{\vec{k'},\lambda^{'}}\right]=(2 \pi)^{2}\de^{(3)}(\vec{k}-\vec{k'})\de_{\lambda,\lambda^{'}},
\eean

\noindent with all other commutators equal to zero, the index $\lambda$ runs over the two polarization states of the graviton and $\epsilon_{ij}$ are the polarization tensors.\\

The equation of motion for the eigenfunctions $\df_{k}(t)$ can be derived in the approximation of de-Sitter space from the second-order action 

\be
S_{2}=\int d\eta^{'}\frac{1}{(H \eta)^{2}}\left[\left(\df^{'}\right)^{2}-\left(\p_i \df \right)^{2}\right],
\ee

\noindent (where $d\eta = dt/a(t)$ is the conformal time) and they are given by the well-known expression

\be\label{uk}
u_{k}(\eta)=\frac{H}{\sqrt{2 k^{3}}}\left(1+i k \eta\right)e^{-ik\eta}.
\ee
In the same approximation, the eigenfunctions for the tensor modes $\gamma_k(\eta)$ are given by $u^{T}_{k}\equiv 2 u_{k}$.\\

Using the positive and negative path technique of the in-in  formalism, the expectation value above can be recast in the form

\be\label{F}
\langle\Omega|\Theta(t)|\Omega\rangle = \left\langle 0\left|T\left(\Theta_{I}(t) e^{-i \int^{t}_{0}dt'\left(H_{I}^{+}(t')-H^{-}_{I}(t')\right)}\right)\right|0\right\rangle,
\ee

\noindent where the plus and minus signs indicate modified Feynman propagators, i.e. modified rules of contraction between interacting fields; schematically we have

\begin{figure}
\begin{center}
\scalebox{0.5}{\includegraphics{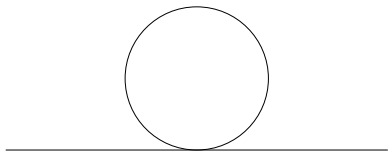}} 
\caption{Diagrammatic representation of the one loop corrections to the power spectrum of $\df$ from scalar modes to leading ($\sim {\epsilon}^{0}$) order in slow-roll.}
\label{1v-scal}
\end{center}
\end{figure}

\begin{figure}
\begin{center}
\scalebox{0.5}{\includegraphics{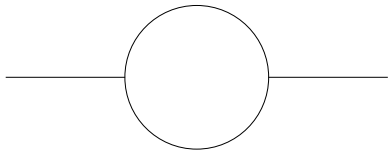}} 
\caption{Next-to-leading ($\sim \sqrt{\epsilon}$) order one loop corrections from scalar modes to the power spectrum of $\df$.}
\label{2v-scal}
\end{center}
\end{figure}

\be
\langle T \left(\phi_{1}\phi_{2}...\phi_{n}\right)\rangle= \sum_{{ij,lm,...}}{[\widehat{\phi_{i}\phi_{j}},\widehat{\phi_{l}\phi_{m}},...]},
\ee

\noindent where the sum is taken over all of the possible sets of field contractions and

\bean
\widehat{\phi^{+}(\epr)\phi^{+}(\eps)} = G^{>}(\epr,\eps)\Theta(\epr-\eps)+G^{<}(\epr,\eps)\Theta(\eps-\epr),\\
\widehat{\phi^{+}(\epr)\phi^{-}(\eps)} = G^{<}(\epr,\eps),\\
\widehat{\phi^{-}(\epr)\phi^{+}(\eps)} = G^{>}(\epr,\eps),\\
\widehat{\phi^{-}(\epr)\phi^{-}(\eps)} = G^{<}(\epr,\eps)\Theta(\epr-\eps)+G^{>}(\epr,\eps)\Theta(\eps-\epr).
\eean

In momentum space we have

\bean
G^{>}_{k}(\epr,\eps)\equiv u_{k}(\epr)u_{k}^{*}(\eps),\\
G^{<}_{k}(\epr,\eps)\equiv u_{k}^{*}(\epr)u_{k}(\eps).
\eean
if $\phi$ is, for example, a neutral scalar field and $u_{k}(\eta)$ its time-dependent wave function. 
Remember that when computing a loop correction the external fields need to be treated like $+$fields.

For our purposes, the exponentials in the expression (\ref{F}) need to be expanded up to second order

\bea \label{3}
\langle\Omega|\Theta(\eta)|\Omega\rangle_{1L}&=&i\Big\langle 0\Big|T\left[\Theta \int^{\eta}_{- \infty}d \epr \left(H_{I}^{+}(\epr)-H_{I}^{-}(\epr)\right)\right]0\rangle\nonumber\\
&+&\frac{(-i)^{2}}{2}\langle 0|T\Big[\Theta \int^{\eta}_{- \infty}d \epr \left(H_{I}^{+}(\epr)-H_{I}^{-}(\epr)\right) \nonumber\\
&\times&\int^{\eta}_{- \infty}d \eps \left(H_{I}^{+}(\eps)-H_{I}^{-}(\eps)\right)\Big]\Big|0\Big\rangle,
\eea 

\noindent with $\Theta(t) \equiv \df_{\vec{k_{1}}}(\eta)\df_{\vec{k_{2}}}(\eta)$ and $H_{I} \equiv H_{I}^{(3)}+H_{I}^{(4)}$, where $H_{I}^{(3)}$ and $H_{I}^{(4)}$ indicate respectively the third and fourth order parts of the interaction hamiltonian. One-loop corrections due to scalars in $H_{I}$ were calculated in \cite{seery1,sloth}; their diagrammatic representation is given in Fig.$1$ for the leading order and in Fig.$2$ for the next-to-leading order corrections. Loop of gravitons were ignored for simplicity in \cite{seery1,sloth}, however they should be included since they are not slow-roll suppressed compared to loops of scalar modes. Their evaluation will constitute the primary goal of this paper. Let us then consider the terms in $H_{I}$ that involve tensor modes. The third order action with gravitons has been calculated in \cite{maldacena}; we will focus on the leading order term in slow-roll parameters and define    

\be
H_{I}^{(3)}(\eta)\equiv \frac{a^{2}(\eta)}{2}\int d^{3}x \ga_{ij}\p_{i}\df\p_{j}\df.
\ee

The fourth order part is given by equation (\ref{Z}). Notice that some of the interaction terms involving the tensor modes in (\ref{Z}) appear with time derivatives. When something like this happens in a theory, the construction of the path integral formula requires additional care compared to the case where time derivatives appear only in the kinetic term of the lagrangian, since extra independent fields (the conjugate momenta) have to be taken into account. This problem will be dealt with in Appendix A. The conclusion of our analysis is that loops that involve conjugate momenta give zero contributions at one loop level to the power spectrum. It is also possible to show that in Eq. (\ref{Z}), of all the leading terms in the slow-roll expansion, only one will provide a non-zero contribution to the loop correction (see Appendix B for a detailed analysis), i.e.

\be
H_{I}^{(4)}(\eta)\equiv  \frac{a^{2}(\eta)}{4}\int d^{3}x\ga_{ik}\ga_{kj}\p_{i}\df\p_{j}\df. 
\ee

\begin{figure}
\begin{center}
\scalebox{0.5}{\includegraphics{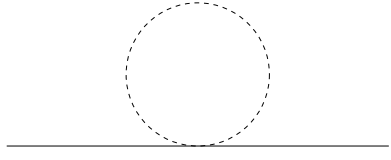}} 
\caption{Diagrammatic representation of the (tensor mode) corrections from $H_{I}^{(4)}$ to the power spectrum of $\df$.}
\label{1v-tens}
\end{center}
\end{figure}

\begin{figure}
\begin{center}
\scalebox{0.5}{\includegraphics{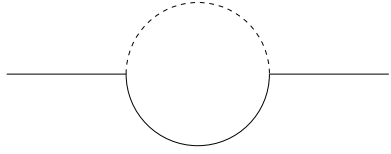}}
\caption{Diagrammatic representation of the (tensor mode) corrections from $H_{I}^{(3)}$ to the power spectrum of $\df$. Notice that this diagram is not slow-roll suppressed compared to the one in Fig.$3$ whereas this is not the case for scalar modes (see Fig.$1$ and Fig.$2$).}
\label{2v-tens}
\end{center}
\end{figure}

Let us now begin with  the one-loop one-vertex part of the diagram (given in Fig.$3$) which we label with the subscript $(1L,1v)$; this can be written as \cite{weinberg1}, \cite{weinberg2}

\be
\langle\df_{\vec{k_{1}}}(\eta^{*})\df_{\vec{k_{2}}}(\eta^{*})\rangle_{(1L,1v)}=i\int d\epr \left\langle\left[H_{I}^{(4)}(\epr),\df_{\vec{k_{1}}}(\eta^{*})\df_{\vec{k_{2}}}(\eta^{*})\right]\right\rangle.
\ee 

We will study this in detail

\bea
\langle\df_{\vec{k_{1}}}(\eta^{*})\df_{\vec{k_{2}}}(\eta^{*})\rangle_{(1L,1v)}=2i\int^{\eta^{*}}_{- \infty} d\epr a^{2}(\epr)\int \frac{d^{3}x}{(2 \pi)^{3}} 
\int d^{3}q_{1}d^{3}q_{2}d^{3}q_{3}d^{3}q_{4}\times\nonumber\\
e^{-i \sum_{n}{\vec{q}}_{n}\cdot\vec{x}}P_{ij}(iq^{i}_{3})(iq^{j}_{4})u_{k_{1}}(\eta^{*})u^{*}_{q_{3}}(\epr)u_{k_{2}}(\eta^{*})u^{*}_{q_{4}}(\epr)u_{q_{1}}(\epr)u^{*}_{q_{2}}(\epr)\times\nonumber\\
\delta^{(3)}(\vec{k_{1}}+\vec{q_{3}})\delta^{(3)}(\vec{k_{2}}+\vec{q_{4}})\delta^{(3)}(\vec{q_{1}}+\vec{q_{2}})+c.c.,
\eea

\noindent where the extra factor of $2$ accounts for the number of equivalent diagrams obtained by permuting the field contractions, $u_{k}(\eta)$ is given by 
Eq.~(\ref{uk}) and

\be P_{ij}k^{i}k^{j}=k^{i}k^{j} \sum_{\lambda,{\lambda}^{'}}\epsilon^{\lambda}_{ik}(\hat{q})\epsilon^{*{\lambda}^{'}}_{kj}(\hat{q})=2 k^{2}\sin^{2}\theta. 
\ee

Integration and the use of the delta function lead to a simpler form

\bea\label{K15}
\fl
\langle\df_{\vec{k_{1}}}(\eta^{*})\df_{\vec{k_{2}}}(\eta^{*})\rangle_{(1L,1v)}&=&-i{\delta}^{(3)}(\vec{k_{1}}+\vec{k_{2}})\frac{H_{*}^{4}}{2k^{4}}\int \frac{d^{3}q}{q^{3}}{\sin}^{2}\theta
\int^{\eta^{*}}_{- \infty}\frac{d {\eta}^{'}}{{\epr}^{2}}e^{2ik(\epr-\eta^{*})}{(1+ik\eta^{*})}^{2} \nonumber \\
&\times& {(1-ik\epr)}^{2}(1+iq\epr)(1-iq\epr)+c.c.,  
\eea
This equation is exact except for the approximation of using the de Sitter space formula for the scale factor, $a(\eta)=-\frac{1}{H\eta}$, and evaluating the Hubble radius $H(\epr)$ at the time $\eta^{*}$. The reason why this is allowed is the following: the contribution to the integral w.r.t. time from regions well before horizon crossing is negligible compared to the contribution due to the region around horizon crossing \cite{maldacena,weinberg1,weinberg2}; in addition to that, we are choosing $\eta$ to be just a few e-folds after horizon crossing, so we can assume that the Hubble radius (as well as any of the slow-roll parameters of the theory) will not undergo a big variation during this interval of time. The same approximation will be applied to the diagrams in the next section.\\
\\
We first solve the time integral. It is convenient to perform a change of variale like in \cite{sloth}, i.e. we set ${x}^{'}=-k\epr$ and $x^{*}=-k\eta^{*}$ 
so that

\bea\label{T}
\fl
\langle\df_{\vec{k_{1}}}(\eta^{*})\df_{\vec{k_{2}}}(\eta^{*})\rangle_{(1L,1v)}&=&{\delta}^{(3)}(\vec{k_{1}}+\vec{k_{2}})\frac{H_{*}^{4}}{2k^{4}}\int \frac{d^{3}q}{q^{3}}{\sin}^{2}\theta\, \, 
Im\Big[\int^{\infty}_{x^{*}}\frac{d{x}^{'}}{k}\frac{k^{2}}{{x}^{'2}}e^{2i({x}^{'}-x^{*})} \nonumber \\
&\times& {(1+ix^{*})}^{2} {(1-i{x}^{'})}^{2}(1+i\frac{q}{k}{x}^{'})(1-i\frac{q}{k}{x}^{'}) \Big]\, .  
\eea
After integrating the imaginary part, we end up with the following result

\bea
\fl
\label{onelooponev}
& & \langle\df_{\vec{k_{1}}}(\eta^{*})\df_{\vec{k_{2}}}(\eta^{*})\rangle_{(1L,1v)}=\\
& & {\delta}^{(3)}(\vec{k_{1}}+\vec{k_{2}})\frac{H_{*}^{4}}{2k^{4}}\int \frac{d^{3}q}{q^{3}}{\sin}^{2}\theta
\frac{2k^{2}(3+x^{*2})+q^{2}(5+5x^{*2}+2x^{*4})}{4k}\nonumber\\
& & ={\delta}^{(3)}(\vec{k_{1}}+\vec{k_{2}})\frac{H_{*}^{4}}{2k^{4}}2\pi\frac{4}{3}
\Big[\frac{k}{2}(3+x^{*2})\int\frac{dq}{q}
+\frac{1}{4k}(5+5x^{*2}+2x^{*4})\int dq q\Big]\, , \nonumber 
\eea

\noindent where the factor $4/3$ comes from integrating with respect to the azimuthal angle $\theta$ (notice that that the reference frame in momentum space has been chosen in such a way that the external wave vector $\vec{k}$ lies along the positive $z$ axis). We now solve the momentum integrals. Both the logarithmic and the quadratic one exhibit ultraviolet divergences and the logarithmic part diverges also at very low momenta. Ultraviolet divergences can be treated as in flat space; the infrared logarithmic divergence is fixed introducing a momentum lower cutoff ${\ell}^{-1}$ to be interpreted as 
a `box  size'~\cite{box4,box5,box1,box2,box3} which can be fixed to be not much larger than the present horizon~\cite{box4,box5}. 
As an example, consider the first integral of Eq.~(\ref{onelooponev}) which is convenient to split as follows  

\be
\int_{{\ell}^{-1}}^{\Lambda}\frac{dq}{q}=\int_{{\ell}^{-1}}^{k}\frac{dq}{q}+\int_{k}^{\Lambda}\frac{dq}{q},
\ee

\noindent where we have introduced an upper cutoff $\Lambda$. The first integral gives $\ln(k \ell)$; the second integral can be renormalized introducing a counterterm $-\ln(\frac{\Lambda}{k_{0}})$, where $k_{0}$ is a renormalization constant. As expected from looking at the diagrammatic representation in Fig. $3$, there is no actual dependence from the external wave number $k$ in the momentum integrals of Eq.~(\ref{onelooponev}). 
The final result for Eq.~(\ref{onelooponev}) can be written as  
\bea\label{FR1}
\fl
\langle\df_{\vec{k_{1}}}(\eta^{*})\df_{\vec{k_{2}}}(\eta^{*})\rangle_{(1L,1v)}=\pi{\delta}^{(3)}(\vec{k_{1}}+\vec{k_{2}})\frac{2H_{*}^{4}}{3k^{3}}(3+x^{*2})\alpha \, ,\\
\nonumber 
\eea
where $\alpha$ is a left over constant from renormalization.\\

Let us now focus on the one-loop contribution from the $3rd$ order action with the gravitons (see Fig.$4$ for its diagrammatic representation)

\bea \label{36}
\fl
\langle\df_{\vec{k_{1}}}(\eta^{*})\df_{\vec{k_{2}}}(\eta^{*})\rangle_{(1L,2v)}&=&\frac{(-i)^{2}}{2}\Big\langle T\Big[\df_{\vec{k_{1}}}(\eta^{*})\df_{\vec{k_{2}}}(\eta^{*}) \int^{\eta^{*}}_{- \infty}d \epr \Big(H_{I}^{+}(\epr)-H_{I}^{-}(\epr)\Big)\nonumber \\
&\times& \int^{\eta^{*}}_{- \infty}d \eps \left(H_{I}^{+}(\eps)-H_{I}^{-}(\eps)\right)\Big]\Big\rangle\nonumber\\
&=& \frac{(-i)^{2}}{2}\Big\langle T\Big[\df_{\vec{k_{1}}}(\eta^{*})\df_{\vec{k_{2}}}(\eta^{*})\Big(A+B+C+D\Big)\Big]\Big\rangle,
\eea 

\noindent where

\bea
A \equiv \int^{\eta^{*}}_{- \infty}d \epr H_{I}^{+}\int^{\eta^{*}}_{- \infty}d \eps H_{I}^{+},\\
B \equiv \int^{\eta^{*}}_{- \infty}d \epr H_{I}^{-}
\int^{\eta^{*}}_{- \infty}d \eps H_{I}^{-},\\
C \equiv -\int^{\eta^{*}}_{- \infty}d \epr H_{I}^{+}\int^{\eta^{*}}_{- \infty}d \eps H_{I}^{-},\\
D \equiv -\int^{\eta^{*}}_{- \infty}d \epr H_{I}^{-}
\int^{\eta^{*}}_{- \infty}d \eps H_{I}^{+}.
\eea

It is easy to check that $B=A^{*}$ and $C=C^{*}=D$. We can write Eq. (\ref{36}) as

\bea 
\langle\df_{\vec{k_{1}}}(\eta^{*})\df_{\vec{k_{2}}}(\eta^{*})\rangle_{(1L,2v)}&=&4 (-i)^{2}\delta^{(3)}(\vec{k_{1}}+\vec{k_{2}})k^{4}\int d^{3}q {\sin}^{4}\theta \\
&\times& \int^{\eta^{*}}_{- \infty}\frac{d \epr}{{(H\epr)}^{2}}\int^{\eta^{*}}_{- \infty}\frac{d \eps}{{(H \eps)}^{2}}\left(w_{f}^{A}+w_{f}^{B}+w_{f}^{C}+w_{f}^{D}\right),\nonumber
\eea

\noindent where the factor $\sin^{4}\theta$ comes from contractions of the polarization tensors with external momenta \cite{polarization}

\be
\epsilon_{ij}(\vec{q})k^{i}k^{j}=\frac{k^{2}}{\sqrt{2}}\left[1-{\left(\frac{\vec{q}\cdot\vec{k}}{qk}\right)}^{2}\right]=\frac{k^{2}}{\sqrt{2}}\sin^{2}\theta,
\ee

\noindent and the wave fuctions $w_{f}$ are

\bea
w_{f}^{A}(\epr,\eps)&=&u_{k}^{}(\eta^{*})u_{k}^{*}(\epr) u_{k}^{}(\eta^{*})u_{k}^{*}(\eps)[u_{|\vec{k}-\vec{q}|}^{}(\epr)u_{|\vec{k}-\vec{q}|}^{*}(\eps)u_{q}^{}(\epr)u_{q}^{*}(\eps)\nonumber\\ 
&\times&\Theta(\epr-\eps)+u_{|\vec{k}-\vec{q}|}^{*}(\epr)u_{|\vec{k}-\vec{q}|}^{}(\eps)u_{q}^{*}(\epr)u_{q}^{}(\eps)\Theta(\eps-\epr)]\, , \nonumber 
\eea
\bea
w_{f}^{B}(\epr,\eps)&=&u_{k}^{*}(\eta^{*})u_{k}^{}(\epr) u_{k}^{*}(\eta^{*})u_{k}^{}(\eps)[u_{|\vec{k}-\vec{q}|}^{}(\epr)u_{|\vec{k}-\vec{q}|}^{*}(\eps)
u_{q}^{}(\epr)u_{q}^{*}(\eps) \nonumber\\
&\times& \Theta(\eps-\epr)
+u_{|\vec{k}-\vec{q}|}^{*}(\epr)u_{|\vec{k}-\vec{q}|}^{}(\eps)u_{q}^{*}(\epr)u_{q}^{}(\eps)\Theta(\epr-\eps)]\, , \nonumber 
\eea
\bea
w_{f}^{C}(\epr,\eps)&=&-u^{}_{k}(\eta^{*})u^{*}_{k}(\epr) u^{*}_{k}(\eta^{*})u^{}_{k}(\eps)u^{*}_{|\vec{k}-\vec{q}|}(\epr)
u^{}_{|\vec{k}-\vec{q}|}(\eps)u^{*}_{q}(\epr)u^{}_{q}(\eps)\, , \nonumber
\eea
\bea
w_{f}^{D}(\epr,\eps)=-u^{*}_{k}(\eta^{*})u^{}_{k}(\epr) u^{}_{k}(\eta^{*})u^{*}_{k}(\eps)u^{}_{|\vec{k}-\vec{q}|}(\epr)
u^{*}_{|\vec{k}-\vec{q}|}(\eps)u^{}_{q}(\epr)u^{*}_{q}(\eps)\, , \nonumber
\eea

\noindent so $w_{f}^{C}(\epr,\eps)=w_{f}^{D}(\eps,\epr)$ and is a real number and $w_{f}^{B}(\epr,\eps)=w_{f}^{A*}(\eps,\epr)$. We will label the two contributions by $A$ and $C$, so that the one-loop contribution with two vertices to the two point function will be broken into two parts

\bea
\fl
\langle\df_{\vec{k_{1}}}(\eta^{*})\df_{\vec{k_{2}}}(\eta^{*})\rangle_{(1L,2v)}&=&
\langle\df_{\vec{k_{1}}}(\eta^{*})\df_{\vec{k_{2}}}(\eta^{*})\rangle_{(1L,2v)}^{A}
+\langle\df_{\vec{k_{1}}}(\eta^{*})\df_{\vec{k_{2}}}(\eta^{*})\rangle_{(1L,2v)}^{C}.
\eea
Let's look in details at the two parts. 

\bea
\fl
\langle\df_{\vec{k_{1}}}(\eta^{*})\df_{\vec{k_{2}}}(\eta^{*})\rangle_{(1L,2v)}^{A}&=&-\delta^{(3)}(\vec{k_{1}}+\vec{k_{2}})\frac{H_{*}^{4}}{2k^{2}}\int\frac{d^{3}q}{q^{3}}\frac{{\sin}^{4}\theta}{{|\vec{k}-\vec{q}|}^{3}}e^{-2ik\eta^{*}}{(1+ik\eta^{*})}^{2} \\
&\times& \int^{\eta^{*}}_{- \infty}\frac{d \epr}{{\epr}^{2}}e^{i\epr(k-q-|\vec{k}-\vec{q}|)}(1-ik\epr)(1+iq\epr)(1+i|\vec{k}-\vec{q}|\epr)
\nonumber\\
&\times& \int^{\epr}_{- \infty}\frac{d \eps}{{\eps}^{2}}e^{i(k+q+|\vec{k}-\vec{q}|)\eps}(1-ik\eps) (1-iq \eps)(1-i|\vec{k}-\vec{q}|\eps)+c.c.\nonumber 
\eea
The second time integral has $e^{i g\eps}\left[-\frac{1}{\eps}+\frac{c}{g}\eps-i\left(\frac{gb-c}{g^{2}}\right)\right]$ as its primitive function, where $g\equiv k+q+|\vec{k}-\vec{q}|$, $b \equiv -qk-(q+k)|\vec{k}-\vec{q}|$ and $c \equiv qk|\vec{k}-\vec{q}|$. This should be evaluated between $-\infty$ and $\epr$. It is soon evident that the lower bound represents a problem for this evaluation. We need to remind ourself, though, that the choice of the integration time contour needs to be deformed and to cross the complex plane to account for the right choice of the vacuum \cite{maldacena}. This is done by integrating in a slightly imaginary direction, i.e. taking $\eps\rightarrow \eps+i\epsilon|\eps|$, where $\epsilon$ is a fixed small real number; so for example

\be
\int_{-\infty}^{\eta} d\epr e^{ik\epr}=\frac{e^{ik\eta}}{i k}.
\ee

With this contour prescription, our integral in $\eps$ vanishes at $- \infty$. Performing the same change of variables as in (\ref{T})

\bea\label{KKKK}
\fl
\langle\df_{\vec{k_{1}}}(\eta^{*})\df_{\vec{k_{2}}}(\eta^{*})\rangle_{(1L,2v)}^{A}&=&-\delta^{(3)}(\vec{k_{1}}+\vec{k_{2}})\frac{H_{*}^{4}}{k^{2}}\int\frac{d^{3}q}{q^{3}}\frac{{\sin}^{4}\theta}{{|\vec{k}-\vec{q}|}^{3}}
Re\Big[\int^{\infty}_{x^{*}}\frac{d{x}^{'}}{k}{\frac{k}{{x}^{'}}}{2}e^{2i({x}^{'}-x^{*})}{(1+ix^{*})}^{2} \nonumber \\
&\times& \left(\frac{k}{{x}^{'}}-\frac{c}{kg}{x}^{'}-i\left(\frac{gb-c}{g^{2}}\right)\right)
\left(1-i\frac{d}{k}{x}^{'}+\frac{s}{k^{2}}{{x}^{'}}^{2}-i\frac{c}{k^{3}}{{x}^{'}}^{3}\right)\Big]\, , \nonumber \\
\eea

\noindent where $d \equiv q-k+|\vec{k}-\vec{q}|$ and $s \equiv kq+(k-q)|\vec{k}-\vec{q}|$. The result of the integration w.r.t.time is a polynomial function of $\sin2x^{*}$, $\cos2x^{*}$, Si($2x^{*}$), Ci($2x^{*}$) and their products with coefficients which depend on $g,b,c,d,s,k$. Notice that in the large scale limit $x^{*}\rightarrow0$ a singulariry similar to the one found in \cite{sloth} shows up in our result. However we evaluate the 
power spectrum of $\delta \phi$ just a few e-folds after horizon crossing, so we are safe from these kind of singular behaviour \cite{seery2}.\\  
The next step consist in performing the momentum integral. The integrals we need to evaluate are of the following kind

\be
\int \frac{d^{3}q}{q^{3}} \frac{{\sin}^{4}\theta}{{|\vec{q}-\vec{k}|}^{3}}f(\vec{q}),
\ee

\noindent where $f(\vec{q})$ is a sum of functions of momentum. Let us begin for simplicity by considering the constant term of the sum, i.e. 
let us study 

\be\label{ex}
\int \frac{d^{3}q}{q^{3}} \frac{{\sin}^{4}\theta}{{|\vec{q}-\vec{k}|}^{3}}.
\ee

For the specific case of equation (\ref{ex}) the integrand function has singularities at $\vec{q}=0$ and at $\vec{q}=\vec{k}$ and shows no ultraviolet singularities. Based on an approximate evaluation performed considering a sphere of radius $\ell^{-1}$ around $\vec{q}=0$, where $\ell^{-1} \ll k$, the integral is proportional to a function $\ln(k\ell)$. The same result can be obtained working in a small sphere around  $\vec{q}=\vec{k}$ after a change of variables $\vec{q_{0}}=\vec{q}-\vec{k}$. The contribution from large values of q is negligible w.r.t. the ones from the singular points, so the integral over the whole momentum space is expected to be proportional to $\ln(k\ell)$. The exact value of the integral can been found 
after a change of variable from the $(q,\theta)$ to the $(q,p)$ space, where $p \equiv |\vec{q}-\vec{k}|$ and is equal to $(16 \pi/225k^{3})\left(1+30\ln(k\ell)\right) \sim k^{-3}\left(10^{-1}+10\ln(k\ell)\right)$.\\

Integrating Eq. (\ref{KKKK}) we find ultaviolet power law and logaritmic singularities in addition to infrared logaritmic contributions. The final result of the integration is a function of $x^{*}=e^{-N_{*}}$, where $N_{*}=\ln(a_{*}/a_{k})$ is the number of e-foldings from horizon crossing
\bea\label{appC1}
\fl
\langle\df_{\vec{k_{1}}}(\eta^{*})\df_{\vec{k_{2}}}(\eta^{*})\rangle_{(1L,2v)}^{A}=\pi\delta^{(3)}(\vec{k_{1}}+\vec{k_{2}})\frac{H_{*}^{4}}{k^{3}}
\Big(a_{1}\ln(k)+a_{2}\ln(k \ell)+a_{3}\Big)\, , \\
\nonumber
\eea
where $a_{1}$, $a_{2}$ and $a_{3}$ are functions of $x^{*}$ (see Appendix C). We are calculating the two point function for the scalar field a few e-foldings after horizon crossing, so $x^{*}$ may be chosen to range between $10^{-1}$ and $10^{-2}$. In this range $a_{1}
\sim {\cal O}(1)$ and negative, $a_{2}=-16/(15 x^{*2})+(8/15)(5-8 
{\rm Ci}(2x^{*}))$ and $a_{3}=-8/(225 x_*^2)+{\cal O}(1)+\rho$, where $\rho$ is  
a left-over scheme-dependent renormalization constant of the kind present in equation (\ref{FR1}). \\

Let us now move to part C of Eq. (\ref{36}) which we give below

\bea
\fl
\langle\df_{\vec{k_{1}}}(\eta^{*})\df_{\vec{k_{2}}}(\eta^{*})\rangle_{(1L,2v)}^{C}&=&\delta^{(3)}(\vec{k_{1}}+\vec{k_{2}})\frac{H_{*}^{4}}{2k^{2}}\int\frac{d^{3}q}{q^{3}}\frac{\sin^{4}\theta}{{|\vec{k}-\vec{q}|}^{3}}\left(1+{(k\eta^{*})^2}\right)
\nonumber\\
&\times& \int^{\eta^{*}}_{- \infty}\frac{d \epr}{{\epr}^{2}}e^{i g \epr}Q(\epr)
\int^{\eta_*}_{- \infty}\frac{d \eps}{{\eps}^{2}}e^{-i g \eps}Q^{*}(\eps)\nonumber\\
&=&\delta^{(3)}(\vec{k_{1}}+\vec{k_{2}})\frac{H_{*}^{4}}{2k^{2}}\int\frac{d^{3}q}{q^{3}}\frac{{\sin}^{4}\theta}{{|\vec{k}-\vec{q}|}^{3}}
\left(1+(k\eta^{*})^2\right)
\nonumber\\
&\times&  \Big[{\left(Re \int d \epr e^{ i g \epr }\frac{Q(\epr)}{{\epr}^{2}}\right)}^{2}+{\left(Im \int d \epr e^{i g \epr }\frac{Q(\epr)}{{\epr}^{2}}\right)}^{2}\Big],
\eea
\noindent where $Q(\epr) \equiv 1+ig\epr+b{\epr}^{2}-ic{\epr}^{3}$.

Let us integrate over conformal time
\bea \label{6}
\fl
\int^{\infty}_{x^{*}}\frac{d{x}^{'}}{k}\frac{k^{2}}{{{x}^{'}}^{2}} e^{-i\frac{g}{k} {x}^{'}}\left[1+i\frac{g}{k}{x}^{'}+\frac{b}{k^{2}}{{x}^{'}}^{2}-i\frac{c}{k^{3}}{{x}^{'}}^{3}\right] =\frac{e^{-i\frac{g}{k}}x^{*}}{k^{2}}\left(-\frac{k^{3}}{x}+\frac{ck}{g}x^{*}+i\frac{(gb-c)k^{2}}{g^{2}}\right)\, , \nonumber \\
\eea

\noindent where again the integration has been performed by continuing $\epr$ to the complex plane, i.e. $(\epr\rightarrow\epr+i\epsilon |\epr|)$, and then taking the limit $\epsilon\rightarrow 0$.\\ 
We are now ready to integrate over momentum

\bea
\fl
\langle\df_{\vec{k_{1}}}(\eta^{*})\df_{\vec{k_{2}}}(\eta^{*})\rangle_{(1L,2v)}^{C}&=&\delta^{(3)}(\vec{k_{1}}
+\vec{k_{2}})\frac{H_{*}^{4}}{2k^{2}}\int\frac{d^{3}q}{q^{3}}\frac{{\sin}^{4}\theta}{{|\vec{k}-\vec{q}|}^{3}}\left(1+{x^{*}}^{2}\right)
\nonumber\\
&\times& \left(\frac{k^{4}}{x^{*2}}-\frac{2k^{2}c}{g}-\frac{2k^{2}bc}{g^{3}}+\frac{k^{2}c^{2}}{g^{4}}+\frac{k^{2}b^{2}}{g^{2}}+\frac{x^{*2}c^{2}}{g^{2}}\right).
\eea

\noindent Similarly to what we have done in part A, one can check that there are no ultraviolet singularities in the remaining five integrals although some infrared logarithmic contributions are still present and the final result is
\bea\label{appC2}
\langle\df_{\vec{k_{1}}}(\eta^{*})\df_{\vec{k_{2}}}(\eta^{*})\rangle_{(1L,2v)}^{C}=\pi\delta^{(3)}(\vec{k_{1}}+\vec{k_{2}})\frac{H_{*}^{4}}{k^{3}}
\left(c_{1}+c_{2}\ln(k \ell)\right)\, , \\
\nonumber
\eea
where $c_{1}=(1/225)\left(8/x^{*2}+107+50x^{*2}\right)$ and $c_{2}=(16/15x^{*2})+(4/15)$. Notice that the $(x^{*})^{-2}$ coefficients in $c_1$ and $c_{2}$ exactly cancels the $(x^{*})^{-2}$ coefficients in $a_{2}$ and $a_3$. This is not surprising: based on \cite{sloth,seery2}, we expect we might observe a logarithmic singularity if we push $x^{*} \rightarrow 0$ in our results (which is indeed present in the Ci(2$x^{*}$) term of $a_{2}$), but no power-law singularities are actually expected.

\section{Final results and conclusions}

Let us now collect our results in the final formula for the power spectrum of the curvature perturbation $\zeta$ computed up to one-loop level. 
This can be derived from Eq.~(\ref{ZZ}), which follows from the $\delta N$ formula, and the final expression reads~\cite{byrnes,seery2}
\bea\label{QQQ}
\langle \zeta_{\vec{k_{1}}}(t)\zeta_{\vec{k_{2}}}(t) \rangle&=&(2 \pi)^{3}\delta^{(3)}(\vec{k_{1}}+\vec{k_{2}})\Big[\left( N^{(1)} \right)^{2}\Big(P_{{\rm tree}}(k_1)
+P_{{\rm one-loop}}(k_1)\Big) \nonumber\\
&+&N^{(1)}N^{(2)}\int \frac{d^{3}q}{(2 \pi)^{3}} B(k_1,q,|\vec{k}_1-\vec{q}|)\nonumber\\
&+&\frac{1}{2}\left( N^{(2)} \right)^{2}\int \frac{d^{3}q}{(2 \pi)^{3}} P_{{\rm tree}}(q)P_{{\rm tree}}(|\vec{k}_1-\vec{q}|)
\nonumber\\
&+&N^{(1)}N^{(3)}P_{{\rm tree}}(k)\int \frac{d^{3}q}{(2 \pi)^{3}} P_{{\rm tree}}(q)\Big]\, ,
\eea
where $P_{{\rm tree}}(k)$ is the tree level power spectrum~(\ref{tree}) and $B(k_{1},k_{2},k_{3})$ is the bispectrum of the scalar field defined by
\bea
\langle {\df}_{\vec{k_{1}}}{\df}_{\vec{k_{2}}}{\df}_{\vec{k_{3}}} \rangle \equiv (2 \pi)^{3}\delta^{(3)}(\vec{k_{1}}+\vec{k_{2}}+\vec{k_{3}})B(k_{1},k_{2},k_{3})\, .
\eea
In previous works the computations of one-loop corrections accounted just for the contributions coming from the inflaton field peturbations (see, for example, 
Refs.~\cite{sloth,sloth2,seery1,seery2}. 
The focus of this paper is on including also the one-loop terms arising from interactions between the tensor (graviton) modes and the scalar field in the case of 
single-field models of slow-roll inflation. At this order the graviton contribution enter the final formula as a correction to the power spectrum of 
the inflaton field around the time of horizon crossing 
\begin{equation} 
P_{{\rm one-loop}}(k)=P_{{\rm scalar}}(k)+P_{{\rm tensor}}(k)\, ,
\end{equation}
and as such it is due to genuine quantum effects, while it does not affect the remaining terms of Eq.~(\ref{QQQ}), which, according to Ref.~\cite{seery2}, 
can be considered as classical contributions arising after the perturbation modes leave the horizon. \footnote{Notice however that 
the computation of $\langle\zeta_{\vec{k_{1}}}(t)\zeta_{\vec{k_{2}}}(t)\rangle$ could be carried out without using the $\delta N$ formalism, working in the comoving gauge and making use only of the Schwinger-Keldysh path integral formula as in \cite{maldacena}. In this respect, the two types of terms, ``classical' and  ``genuine quantum effects'', are not really supposed to be entirely different.}  

Summing the main results of the previous section, Eqs. (\ref{FR1}), (\ref{appC1}) and (\ref{appC2}), we finally find the one-loop graviton correction to the inflaton power spectrum 

\be\label{FR}
\langle\df_{\vec{k_{1}}}(\eta^{*})\df_{\vec{k_{2}}}(\eta^{*})\rangle_{1L}=\pi\delta^{(3)}(\vec{k_{1}}+\vec{k_{2}})\frac{H_{*}^{4}}{k^{3}}
\left[f_{1}\ln(k)+f_{2}\ln(k \ell)+f_{3}\right]\, ,
\ee
where 
\bea
f_{1}=-\frac{4}{15}\left(5+5x^{*2}+2x^{*4}\right)\, ,
\eea
\bea
f_{2} =a_{2}+c_{2}\, ,
\eea
and $f_{3}$ is given by a left-over scheme-dependent renormalization constant plus contributions of order 
${\cal O}(1)$ (see Appendix C for the complete expressions of $a_{2}$, $c_{2}$ and $f_3$). If we calculate the two point funtion of $\df$ a few e-foldings after horizon crossing, i.e. $x^{*}$ ranges for example between $10^{-1}$ and $10^{-2}$, $f_{1}$ reduces to a negative constant of order ${\cal O}(1)$ and $f_{2}\sim 4\left(1-{\rm Ci}(2x^{*})\right) \sim {\cal O}(10)$. In the limit where 
$x^{*} \rightarrow 1$ both $f_{1}$ and $f_2$ turn out to be of order unity.   

In order to understand which is the dominant contribution in Eq.~(\ref{QQQ}) and how big it is, one needs to 
(i) know the slow-roll order of the coefficients $N^{(i)}$: $N^{(1)} \sim {\epsilon}^{-1/2}$, $N^{(2)} \sim \epsilon^{0}$, 
$N^{(3)} \sim \epsilon^{1/2}$; 
(ii) compute the integrals involving the power spetrum $P(q)$. 
This is discussed in details in Ref.~\cite{seery2} (for the case of scalar perturbations only), see in particular Sec IV of~\cite{seery2}. 
It turns out that the crucial quantity is represented by the number of 
e-foldings of inflation between the times of 
horizon exit of the mode $\ell^{-1}$, which corresponds to the infrared cutoff, and the time of horizon exit of the mode $k$ we want to observe. 
However, to deal with observable quantitites one has to choose $\ell$ not much bigger than the present cosmological horizon $H_0^{-1}$~\cite{box4,box5}.\\
The relevant point about Eq. (\ref{FR}) is that it gives in Eq.~(\ref{QQQ}) a contribution which is of the same order of magnitude as 
those coming from loops which accounts for scalar perturbations only. Since in terms of the slow-roll parameters 
$\left( N^{(1)} \right)^2 \sim {\epsilon_*}^{-1}$ the magnitude of the one-loop graviton correction turns out to be
\begin{equation}
\label{fin}
\Delta P_{\zeta}^{1\rm{loop}}(k) \sim \frac{2 \pi^2}{k^3} \alpha(k) \frac{1}{\epsilon_*}{\cal P}^2_*(k)\, , 
\end{equation}  
where we have used Eq.~(\ref{tree}) for the power spectrum of the inflaton field. In Eq.~(\ref{fin}) 
$\alpha(k)$ includes the various coefficients of Eq.~(\ref{FR}), and it is ${\cal O}(1)$. 
Eq.~(\ref{fin}) 
allows a more direct comparison with the results of Ref.~\cite{seery2}, showing that the graviton contributions to the one-loop 
corrections are comparable to the ones computed only from scalar interactions. Notice that also for the tensor contributions we find terms of the form 
$\ln(k\ell)$. 

One-loop corrections to the power spectrum of density perturbations from inflation are 
small, still a precise and self-consistent computation requires to account also for the loops which are switched on by the tensor (graviton) modes. 
First, going beyond linear order, the tensor perturbation modes produced during inflation unavoidably mix with scalar modes. 
This fact alone would require to include 
the tensor modes for a self-consistent computation. 
Most importantly, despite a naive expectation suggested by the fact that the power spectrum of the tensor modes is suppressed 
on large scales with respect to that of the curvature (scalar) perturbations, 
our results show explicitly that their inclusion is necessary since their contribution is not at all negligible with respect to the
loop corrections arising from interactions involving the inflaton field only.

\section*{Acknowledgments}
We would like to thank Peter\,Adshead, Richard\,Easther, Eugene\,Lim, Sabino\,Matarrese, Massimo\,Pietroni, Antonio\,Riotto, 
David\,Seery, Martin\,Sloth and Alberto\,Vallinotto, 
for important discussions and correspondence. This research has been partially supported by ASI contract I/016/07/0
"COFIS" and ASI contract Planck LFI Activity of Phase E2.

\section*{Appendix A. Path integral formalism for a lagrangian with time-dependent interactions}

The propagator of two fields $\phi_{1}$ and $\phi_{2}$ is defined by (see for example \cite{peskin})

\be
\langle\phi_{1}\phi_{2}\rangle=\int D\phi D\Pi e^{i\int d^{4}x\left(\Pi\dot{\phi}-\textit{H}\right)},
\ee

\noindent where $\Pi$ is the momentum conjugate to $\phi$ and $\textit{H}$ is the hamiltonian density. If $H$ is quadratic in $\Pi$, as it happens for example in flat space-time for a field governed by a lagrangian $L=\int d^{4}x \left(\frac{1}{2}\p_{\mu}\p^{\mu}\phi-V(\phi)\right)$ the square in the exponent can be completed and the integral in $\Pi$ evaluated and all is left is

\be
\langle\phi_{1}\phi_{2}\rangle=\int D\phi e^{iL}.
\ee
 
So, if interaction with time derivatives appear in the lagrangian, $\Pi$ and $\phi$ are independent fields in the path integral. This will provide some extra vertices that need to be accounted for in the Feynman diagrams. We sketch a derivation of these extra vertices. It will turn out to be similar to what Seery does in \cite{seery1}, although complicated by the presence of gravitons. To keep the calculations easier we will at first ignore spatial derivatives and tensor indices, this will also make the notation simpler. Also, we won't keep track of all of the numerical real coefficients since this is not crucial for getting to the final result; it is instead very important to keep track of imaginary coefficients, time derivatives and powers of the scale factor $a$ factors, and we will make sure they are all accounted for in our analysis.\\
The total action is $S=\int d\eta \left(L_{\gamma}+L_{\phi}\right)$, where

\bea
L_{\gamma}&=&a^{2}{\gap}^{2}+\Ga_{\ga}\gap {\dfp}^{2}+\Ga_{\ga}{\gap}^{3}+\Ga_{\phi}\dfp{\gap}^{2}+\Ga_{\ga \phi}\dfp\gap+\Ga_{\ga \ga}{\gap}^{2}\nonumber\\&+&\lambda_{\phi \ga \ga}\dfp
+\lambda_{\phi \phi \ga}\gap\nonumber+\lambda_{\ga \ga \ga}\gap,\\
L_{\phi}&=&a^{2}{\dfp}^{2}+\Ga_{1}{\dfp}^{2}+\Ga_{2}{\dfp}^{2}+\omega {\dfp}^{3}+\lambda \dfp,
\eea

\noindent where $f^{'} \equiv df / d\eta$ and where we define

\bea
\Ga_{\ga}\sim a \ga\label{K1},\\
\Ga_{\phi}\sim a \df ,\\
\Ga_{1}\sim \dot{\phi} a^{2}\df,\\
\Ga_{2}\sim a^{2}{\df}^{2},\\
\Ga_{\ga \phi}\sim a^{2}\df \ga,\\
\Ga_{\ga \ga}\sim a^{2} {\ga}^{2},\\
\omega \sim a \df,\\
\lambda_{\phi \phi \phi} \sim a {\df}^{3},\\
\lambda_{\phi \ga \ga}\sim a \df {\ga}^{2},\\ 
\lambda_{\phi \phi \ga}\sim a \ga {\df}^{2},\\
\lambda_{\ga \ga \ga}\sim a {\ga}^{3}.\label{K2}
\eea

Notice that in equations (\ref{K1}) through (\ref{K2}) we use the equivalence symbol meaning that we skeep details about integrations in momenta and real coefficients.\\

The conjugate momenta are

\bea
\fl
\Pig\equiv \frac{\delta L}{\delta \gap}= a^{2}\gap+\lambda_{\ga \ga \ga}+\lambda_{\phi \phi \ga}+\Ga_{\ga \phi}\dfp+\Ga_{\ga}{\dfp}^{2}  
+\Ga_{\phi}\gap\dfp+\Ga_{\ga \ga}{\gap}^{3}+\Ga_{\ga}{\gap}^{2}\label{K3},\\
\nonumber\\
\fl
\Pif\equiv \frac{\delta L}{\delta \left(\dfp\right)}=\Ga_{\ga}\gap\dfp+\Ga_{\ga \phi}\gap+\lambda_{\phi \phi \ga}+a^{2}\dfp+\Ga_{1}\dfp+\Ga_{2}\dfp
+\lambda_{\phi \phi \phi}+\omega{\dfp}^{2}.\label{K4}
\eea
We solve perturbatively the equations (\ref{K3}) and (\ref{K4}) in order to derive $\gap$ and $\dfp$ to fourth order

\bea
\gap&=&a^{-2}\Big[\Pig+\lambda_{\ga \ga \ga}+\lambda_{\phi \phi \ga}+a^{-2}\Ga_{\ga \phi}\Pif+a^{-4}\Ga_{\ga}\Pif\Pif+a^{-4}\Ga_{\phi}\Pig\Pif\nonumber\\&+&a^{-2}\Ga_{\ga \ga}\Pig+a^{-4}\Ga_{\ga}\Pig\Pig
+a^{-4}\Ga_{\ga \phi}\Ga_{1}\Pif+a^{-4}\Ga_{\phi}\Pig\Pif\nonumber\\
&+&a^{-6}\Ga_{\phi}\Ga_{1}\Pig\Pif+a^{-2}\Ga_{\ga \ga}\Pif\Big],\label{K5}\\
\dfp&=&a^{-2}\Big[\Pif+a^{-4}\Ga_{\ga}\Pig\Pif+a^{-6}\Ga_{\ga}\Ga_{1}\Pig\Pif+a^{-2}\Ga_{\ga \phi}\Pig+\lambda_{\phi \ga \ga}\nonumber\\
&+&a^{-2}\Ga_{1}\Pif+a^{-4\Ga_{1}}\Ga_{1}\Pif+a^{-2}\Ga_{2}\Pif+\lambda_{\phi \phi \phi}+a^{-4}\omega\Pif\Pif\nonumber\\&+&a^{-4}\Ga_{\ga \phi}\Ga_{1}\Pig+a^{-2}\Ga_{1}\lambda_{\phi \ga \ga}+a^{-6}\Ga_{1}\Ga_{1}\Ga_{1}\Pif+a^{-4}\Ga_{1}\Ga_{2}\Pif\nonumber\\&+&a^{-2}\Ga_{1}\lambda_{\phi \phi \phi}+a^{-6}\Ga_{1}\omega\Pif\Pif\Big].\label{K6}
\eea

The next steps are: derive the hamiltonian $H=\Pig \gap+\Pif \dfp-L\left(\ga,\df,\gap,\dfp\right)$, where we need to plug in the solution (\ref{K5}) and (\ref{K6}) for $\gap$ and $\dfp$; construct the action as $S=S_{0}+S_{\Pi}$, where $S_{0}=\int d \eta \left(L_{\ga}+L_{\phi}\right)$ and $S_{\Pi}$ includes the terms that depend on the conjugate momenta of the fields (a change of variables similar to the one that Seery performs in \cite{seery1} over the conjugate momenta will also be necessary).\\ 
Let's consider the vertices in $S_{\Pi}$ that are involved in the corrections to the one loop point function for the scalar field

\bea\label{PI}
S_{\Pi}&\supset& \int d\epr \Big[a^{-4}\Ga_{1}\dfp\Pif+a^{-4}\Ga_{1}\Pif\Pif+a^{-2}\Ga_{1}\gap\Pif+a^{-4}\Ga_{2}\Pif\Pif\nonumber\\
&+&a^{-4}\Ga_{\phi}\dfp\Pig\Pig+a^{-4}\omega\dfp\Pif\Pif\Big].\label{K10}
\eea   

The first three vertices belong to the third order part of the action; $a^{-4}\Ga_{1}\dfp\Pif$ and $a^{-2}\Ga_{1}\gap\Pif$ provide a correction to the two point function at one loop with two vertices. Because of the presence of $\Ga_{1}$ which involves a factor of $\dot{\phi}$, it is subleading in slow roll order w.r.t. the corrections coming from fourth order vertices. We will therefore neglect these diagrams. The same applies to the second vertex, $a^{-4}\Ga_{1}\Pif\Pif$, although this may contribute to correcting the one point function

\be\label{K8}
\langle\df_{\vec{k}}(\eta^{*})\rangle \supset C_{1} \frac{H_{*}^{2}}{k^{3}}\sqrt{\epsilon}\int d^{3}q f_{1}(\vec{q}) \int_{- \infty}^{\eta^{*}}d\epr \delta(0) \left(1-ik\epr\right)e^{ik\epr}+c.c.,
\ee

\noindent where $C_{1}$ is a numerical real coefficient and $\delta(0)$ is the Dirac delta function deriving from the propagator of the $\Pif$'s and $f_{1}$ is a scalar function of the internal momentum. The main contribution to the integral is due to times around horizon crossing since at early times the rotation to imaginary plane of the contour integral makes the exponent decrease rapidly to zero and moreover $\eta^{*}$ was chosen to be just a few e-folding after horizon crossing. Also, since the integrand function goes to zero as $\epr$ approaches zero, we get a good approximation of this integral taking the upper limit $\eta^{*}\rightarrow 0$. The result is purely imaginary and it cancels out with its complex conjugate.\\
Let's now move to the fourth order vertices (Fig.$6$). From $a^{-4}\Ga_{2}\Pif\Pif$ we have

\be
\langle\df_{\vec{k_{1}}}(\eta^{*})\df_{\vec{k_{2}}}(\eta^{*})\rangle \supset \frac{H_{*}^{4}}{k^{6}} C_{2} \int d^{3}q f_{2}(\vec{q})\int^{\eta^{*}}_{-\infty}d \epr\delta(0){\left(1-ik\epr\right)}^{2}e^{2ik\epr}+c.c.
\ee   

\noindent The same consideration as in (\ref{K8}) apply to the integral above, which gives a zero contribution, as well as the following diagrams (corresponding to the last two vertices in (\ref{K10})) 

\bea
\fl
\langle\df_{\vec{k_{1}}}(\eta^{*})\df_{\vec{k_{2}}}(\eta^{*})\rangle &\supset& \frac{H_{*}^{4}}{k^{6}} C_{3} \int d^{3}q f_{3}(\vec{q})\int^{\eta^{*}}_{-\infty}d\epr\delta(0){\epr}^{2}\left(1-ik\epr\right)e^{2ik\epr}\nonumber\\
&+&\frac{H^{4}}{k^{6}} C_{4} \int d^{3}q f_{4}(\vec{q})\int^{\eta^{*}}_{-\infty}d \epr\delta(0){\epr}^{2}\left(1-ik\epr\right)e^{2ik\epr}+c.c.
\eea

\section*{Appendix B. Study of leading slow roll order vertices in the fourth order action}

We are interested in computing the correlators just a few e-foldings after the scales we consider cross the horizon, so we can assume that the slow roll parameters remain small and can be treated as constants during this length of time. It is then correct to limit our interest to the leading order slow-roll contribution to the action .\\

Let's start from the study of the slow-roll order of the fluctuations derived as solution to the constraint equations:

\bean
\alo &=& \sqrt{\ep} \textit{Q}_{\alo}[\df],\\
\tho &=& \sqrt{\ep} \textit{Q}_{\tho}[\df],\\
\alt &=& \ep\textit{R}_{\alt}[\df^{2}]+\sqrt{\ep}\textit{S}_{\alt}[\df,\ga]+\textit{T}_{\alt}[\df^{2}],\\
\tht&=&\ep\textit{R}_{\tht}[\df^{2}]+\sqrt{\ep}\textit{S}_{\tht}[\df,\ga]+\textit{T}_{\tht}[\df^{2}]+\ep^{2}\textit{U}_{\tht}[\df^{2}]+\textit{V}_{\tht}[\ga^{2}]
\nonumber\\&+&\ep^{3/2}\textit{W}_{\tht}[\df,\ga],\\ \b_{j}&=&\ep\textit{R}_{j}[\df^{2}]+\sqrt{\ep}\textit{S}_{j}[\df,\ga]+\textit{T}_{j}[\df^{2}]+\textit{V}_{j}[\ga^{2}],
\eean

\noindent where $\textit{S}[\df,\ga]$ is a linear function of $\df$ and/or its derivatives and a linear function of $\ga$ and/or its derivatives, $\textit{R}[\df^{2}]$ is a quadratic function of $\df$ and/or its derivatives and so on.  

Notice that the first order fluctuations are subleading ($\sim \sqrt{\ep}$) w.r.t. the second order ones ($\sim \ep^{0}$). This criterium allows a suppression of a large number of terms in the $4th$ order action based on keeping the leading order (i.e. $\sim \ep^{0}$) terms only

\bean
S_{4}&=&a^{3} \int dt d^{3}x\Big[\frac{1}{4a^{2}}\Big(\p_{i}\b_{j}+\p_{j}\b_{i}\Big)\p_{i}\b_{j}
-\frac{1}{a^{2}}\ddf\left(\p_{j}\th_{2}+\b_{j}\right)\p_{j}\df+3H^{2}\al_{2}^{2}\\
&-&\frac{1}{a^{2}}\left(\frac{1}{4}\ga_{ik}\ga_{kj}\p_{j}\df\p_{i}\df+\alt\p_{i}\df\p_{i}\df-\p_{k}\tht\dot{\ga_{ab}}\p_{b}\ga_{ak}+\frac{1}{2}\dot{\ga_{ab}}\p_{k}\ga_{ab}\p_{k}\tht\right)\\
&+&\frac{1}{a^{2}}\left(\b_{k}\dot{\ga_{ab}}\p_{b}\ga_{ak}-\frac{1}{2}\dot{\ga_{ab}}\b_{k}\p_{k}\ga_{ab}\right)\Big].
\eean

\noindent It can be easily shown that the terms in the action that do not contain the gravitons reproduce the ones in equation (37) of \cite{sloth-seery-lidsey}. The contribution to the power spectrum due to these vertices has been calculated by these authors, but only for the scalar part. We will then focus on all the tensor contributions from these and from the remaining terms. Interaction vertices with both two and four tensor fluctuations will be obtained once the expressions for $\alt$, $\tht$ and $\b_{j}$ are plugged in the action. The terms in the action that we need for constructing Feynman diagrams with one loop of gravitons are

\bea\label{Z}
S_{\ga^{2}}&=&a^{3}\int d^{3}x dt \Big[-\frac{1}{4a^{2}}\b_{j}\p^{2}\b_{j}-\frac{1}{a^{2}}\ddf\p_{j}\df\p_{j}\tht-\frac{1}{4 a^{2}}\p_{j}\df\p_{i}\df\ga_{ik}\ga_{kj}\\&-&\frac{1}{a^{2}}\ddf\p_{j}\df\b_{j}
+\frac{1}{2a^{2}}\Big(2\dot{\ga_{ab}}\p_{a}\ga_{ak}\left(\p_{k}\tht+\b_{k}\right)-\dot{\ga_{ab}}\p_{k}\ga_{ab}\Big(\b_{k}+\p_{k}\tht\Big)\Big]\nonumber.
\eea

Let's plug the expressions for $\beta_{j}$ and $\tht$ into (\ref{Z}) considering the terms with two gravitons. The result is an ensemble of vertices which can in principle contribute to the one loop corrections to the power spectrum of the scalar field. Apart from $\p_{j}\df\p_{i}\df\ga_{ik}\ga_{kj}$, all of the other terms contain time derivatives of one, two or three of the four fields

\bea
\b_{j}\p^{2}\b_{j}&\supset&  a^{4}\Big[\p^{-4}\Big(\p_{m}\p_{j}\ddf\p_{m}\df-\p^{2}\ddf\p_{j}\df+\p_{j}\ddf\p^{2}\df\nonumber\\&-&\p_{m}\ddf\p_{m}\p_{j}\df\Big)\left(\dot{\ga_{ik}}\p_{i}\ga_{kj}-\ga_{il}\p_{i}\dot{\ga_{kj}}\right)\nonumber\\&+&\p^{-2}\Big(\p_{m}\p_{j}\ddf\p_{m}\df-\p^{2}\ddf\p_{j}\df+\p_{j}\ddf\p^{2}\df\nonumber\\&-&\p_{m}\ddf\p_{m}\p_{j}\df\Big)\p^{-2}\left(\dot{\ga_{ik}}\p_{i}\ga_{kj}-\ga_{il}\p_{i}\dot{\ga_{kj}}\right)\Big],
\eea
\bea
\ddf\p_{j}\df\p_{j}\tht &\supset& \ddf\p_{j}\df\frac{1}{16H}\p^{-2}\p_{j}\Big[\frac{1}{2a^{2}}\p_{a}\ga_{iq}\p_{a}\ga_{iq}+\dot{\ga_{lj}}\dot{\ga_{lj}}\Big]
\eea
\bea
\ddf\p_{j}\df\b_{j}&\supset& \ddf\p_{j}\df\frac{a^{2}}{2}\p^{-2}\p_{j}\Big[\dot{\ga_{ik}}\p_{i}\ga_{kj}-\ga_{ik}\p_{i}\dot{\ga_{kj}}\Big],
\eea
\bea
\dot{\ga_{ab}}\p_{a}\ga_{bk}\p_{k}\tht&\supset&\dot{\ga_{ab}}\p_{a}\ga_{bk}\frac{a^{2}}{4H}\p^{-2}\p_{k}\Big[-6H\p^{-2}\Sigma-{\ddf}^{2}-\frac{1}{a^{2}}\p_{i}\df\p_{i}\df\Big],
\eea
\bea
\dot{\ga_{ab}}\p_{a}\ga_{bk}\b_{k}&\supset&\dot{\ga_{ab}}\p_{a}\ga_{bk}2a^{2}\p^{-4}\Big[\p_{m}\p_{k}\ddf\p_{m}\df-\p^{2}\ddf\p_{k}\df\nonumber\\&+&\p_{k}\ddf\p^{2}\df-\p_{m}\ddf\p_{m}\p_{k}\df\Big],
\eea
\bea
\dot{\ga_{ab}}\p_{k}\ga_{ab}\beta_{k}&\supset& \dot{\ga_{ab}}\p_{k}\ga_{ab}\p^{-4}\Big[\p_{m}\p_{k}\ddf \p_{m}\df-\p^{2}\ddf\p_{k}\df+\p_{k}\ddf\p^{2}\df\nonumber\\&-&\p_{m}\ddf\p_{m}\p_{k}\df\Big]\label{V3ref1}
\eea
\bea
\dot{\ga_{ab}}\p_{k}\ga_{ab}\p_{k}\tht&\supset& \dot{\ga_{ab}}\p_{k}\ga_{ab}\Big[-6H\p^{-2}\Big(\p^{2}\df\ddf+\p_{j}\df\p_{j}\ddf\Big)\nonumber\\&-&{\ddf}^{2}-\frac{1}{a^{2}}\p_{i}\df\p_{i}\df\Big] \label{V3ref2}.
\eea

We will now prove that the vertices that include time derivatives do not actually contribute to the two point function. First of all notice that the tensor fields carry polarization tensors $\epsilon_{ij}$ with the property $q^{i}\epsilon_{ij}=0$ and are always contracted with other tensor fields in the calculations; this implies that, if a partial derivative index is contracted with a tensor index, that diagram will be zero. Based on this observation, we can ignore several of the vertices with time derivatives. We are eventually left with only two of them, that we will call $V_{1}$, $V_{2}$ and $V_{3}$ 

\bea
V_{1} \sim \p_{j}\left(\ddf\p_{j}\df\right)\p^{-2}\left(\p_{a}\ga_{bc}\p_{a}\ga_{bc}\right),\\
\nonumber\\
V_{2} \sim \p_{j}\left(\ddf\p_{j}\df\right)\p^{-2}\left(\dot{\ga_{ab}}\dot{\ga_{ab}}\right),\\
\nonumber\\
V_{3} \sim \dot{\ga_{ab}}\p_{k}\ga_{ab}\Big(\beta_{k}+\p_{k}\tht\Big)\label{V3}.
\eea

\noindent where (\ref{V3}) is given by the sum of (\ref{V3ref1}) and (\ref{V3ref2}). Notice that the $\ga_{ij}$ fields need to be contracted between each other and that $\sum_{\lambda,{\lambda}^{'}}\epsilon^{\lambda*}_{iq}\epsilon^{{\lambda}^{'}}_{iq}=$constant \cite{polarization}; the derivatives of $\df$ contract with derivatives of $\ga$, so this produces $\vec{k} \cdot \vec{q}$ factors. Therefore we have

\bea
\fl
\langle\df_{\vec{k_{1}}}(\eta^{*})\df_{\vec{k_{2}}}(\eta^{*})\rangle_{V_{1}+V_{2}} \sim i \delta^{(3)}(\vec{k_{1}}+\vec{k_{2}})H_{*}^{4}\int \frac{d^{3}q}{q^{3}}f_{1}(q^{2})\vec{k} \cdot \vec{q}\int^{\eta^{*}}_{- \infty}d \epr f_{2}(\epr)+c.c.,   
\eea

\noindent where $f_{1}(q^{2})$ and $f_{2}(\epr)$ are some functions of $q^{2}$ and $\epr$. This contribution is evidently zero for symmetry reasons.\\

\section*{Appendix C. Complete expressions of one-loop two-vertex diagrams to leading order }
In the following we provide the explicit expression for Eqs.~(\ref{appC1}) and (\ref{appC2}). 
Equation (\ref{appC1}) reads as

\bea
\fl
\langle\df_{\vec{k_{1}}}(\eta^{*})\df_{\vec{k_{2}}}(\eta^{*})\rangle_{(1L,2v)}^{A}=\pi\delta^{(3)}(\vec{k_{1}}+\vec{k_{2}})\frac{H_{*}^{4}}{k^{3}}
\Big(a_{1}\ln(k)+a_{2}\ln(k \ell)+a_{3}\Big),
\eea
where

\bea
a_{1} &=& -\frac{4}{15}\Big(5+5x^{*2}+2x^{*4}\Big),
\eea

\bea
a_{2} &=& \frac{8}{15x^{*2}}\Big[-2
+\Big(5-8\sigma_{c}\tilde{\sigma}_{c}+4\pi \sigma_{s}-8\sigma_{s}\tilde{\sigma}_{s}\Big)x^{*2}-8(\pi \sigma_{c}+2\tilde{\sigma}_{c}\sigma_{s}\nonumber\\&-&2\sigma_{c}\tilde{\sigma}_{c})x^{*3}
+(1+8\sigma_{c}\tilde{\sigma}_{c}-4 \pi \sigma_{s}+8\sigma_{s}\tilde{\sigma}_{s})x^{*4}\Big],
\eea

\bea
a_{3} &=& \frac{1}{1800x^{*2}}\Big[-64
-\Big(-3120+15136\sigma_{c}\tilde{\sigma}_{c}
-450{\pi}^{2}\sigma_{c}\tilde{\sigma}_{c}-7568\pi \sigma_{s}\nonumber\\&+&225{\pi}^{3}\sigma_{s}+15136\sigma_{s}\tilde{\sigma}_{s}\Big)x^{*2}
-\Big(15136\pi \sigma_{c}-450{\pi}^{3}\sigma_{c}+30272\tilde{\sigma}_{c}\sigma_{s}\nonumber\\&-&900{\pi}^{2}\tilde{\sigma}_{c}\sigma_{s}
-30272\sigma_{c}\tilde{\sigma}_{s}+900{\pi}^{2}\sigma_{c}\tilde{\sigma}_{s}\Big)x^{*3}
-\Big(-672-15136\sigma_{c}\tilde{\sigma}_{c}\nonumber\\
&+&450{\pi}^{2}\sigma_{c}\tilde{\sigma}_{c}+7568\pi \sigma_{s}
-225{\pi}^{3}\sigma_{s}
-15136\sigma_{s}\tilde{\sigma}_{c}+450{\pi}^{2}\sigma_{s}\tilde{\sigma}_{s}\Big)x^{*4}\nonumber\\
&-&208x^{*6}\Big]+\rho,
\eea
and $\rho$ is a constant left over from renormalization of ultraviolet divergences. We have defined

\bea
\sigma_{s} \equiv \sin2x^{*},\nonumber\\
\sigma_{c} \equiv \cos2x^{*},\nonumber\\ 
\tilde{\sigma}_{s} \equiv {\rm Si}(2x^{*}),\nonumber\\
\tilde{\sigma}_{c} \equiv {\rm Ci}(2x^{*}),\nonumber
\eea

\noindent where ${\rm Si}$ and ${\rm Ci}$ stand for the sine-integral and the cosine-integral functions, i.e. 

\bea
{\rm Si}(x)=\int^{x}_{0}\frac{\sin(t)}{t}dt,\nonumber\\
{\rm Ci}(x)=\int^{x}_{0}\frac{\cos(t)-1}{t}dt+\ln(x)+\gamma,\nonumber
\eea

\noindent with $\gamma$ indicating the Euler Gamma function.\\
The expression for Eq.~(\ref{appC2}) is

\bea
\langle\df_{\vec{k_{1}}}(\eta^{*})\df_{\vec{k_{2}}}(\eta^{*})\rangle_{(1L,2v)}^{C}=\pi\delta^{(3)}(\vec{k_{1}}+\vec{k_{2}})\frac{H_{*}^{4}}{k^{3}}
\left(c_{1}+c_{2}\ln(k \ell)\right),
\eea
where

\bea
c_{1}&=& \frac{1}{225}\left(\frac{8}{x^{*2}}+107+50x^{*2}\right),\\
\nonumber\\
c_{2}&=& \frac{4}{15}\left(\frac{4}{x^{*2}}+1\right).
\eea

Finally the quantity $f_{3}$ appearing in Eq.~(\ref{FR}) is given by

\be
f_{3}=a_{3}+c_{1}+{\alpha}^{'},
\ee
where ${\alpha}^{'}\equiv 2\left(1+x^{*2}/3\right)\alpha$ from Eq.~(\ref{FR1}).

\end{document}